\documentclass[12pt]{article} 
\usepackage[hyperfootnotes=false]{hyperref}
\usepackage{amsmath}
\usepackage{amssymb}
\usepackage{graphicx}
\usepackage{xcolor}
\newlength{\abstractwidth}
\usepackage{mciteplus} 
\usepackage{subcaption}
\usepackage{bbold}
\usepackage{slashed}

\newcommand{\be}{\begin{equation}}
\newcommand{\bea}{\begin{eqnarray}}
\newcommand{\eea}{\end{eqnarray}}
\newcommand{\beq}{\begin{equation}}
\newcommand{\ee}{\end{equation}}
\newcommand{\eeq}{\end{equation}}

\newcommand{\half}{{1\over 2}}

\def \lll {\left(}
\def \rrr {\right)}
\def \pr {\partial}

\def \th {\theta}

\def\la{\label}

\def\32{{3 \over 2 } }

\def\ba{\begin{eqnarray}}
\def\ea{\end{eqnarray}}

\def\simleq{\; \raise0.3ex\hbox{$<$\kern-0.75em
      \raise-1.1ex\hbox{$\sim$}}\; }
\def\simgeq{\; \raise0.3ex\hbox{$>$\kern-0.75em
      \raise-1.1ex\hbox{$\sim$}}\; }

\def\nref#1{(\ref{#1})}

\begin{document}

\begin{titlepage}
  \bigskip

  \bigskip\bigskip

  \bigskip

\begin{center}
 
\centerline
{\Large \bf {Traversable wormholes in four dimensions}}
 \bigskip

 \bigskip
{\Large \bf { }} 
    \bigskip
\bigskip
\end{center}

  \begin{center}

 \bf {Juan Maldacena$^1$,  Alexey Milekhin$^{2}$ and Fedor Popov$^2$   }
  \bigskip \rm
  
\bigskip
 $^1$Institute for Advanced Study,  Princeton, NJ 08540, U.S.A.  

 \rm 
 \bigskip
   $^2$Physics Department, Princeton University, Princeton, NJ 08544, U.S.A.\\
\rm
 \bigskip

  \bigskip \rm
\bigskip
 
\rm

\bigskip
\bigskip

  \end{center}

 \bigskip\bigskip
  \begin{abstract}

We present a wormhole solution in four dimensions. It is a solution of an Einstein Maxwell theory plus charged massless fermions. The fermions give rise to a negative Casimir-like energy, which makes the wormhole possible. 
  It is a long wormhole that does not lead to causality violations in the ambient space. 
  It can be viewed as a pair of entangled near extremal black holes with an interaction term generated by the exchange of fermion fields. 
  The solution can be embedded in the Standard Model by making its overall size  small   compared to the electroweak scale.

 \medskip
  \noindent
  \end{abstract}
\bigskip \bigskip \bigskip

\vspace{1cm}

\vspace{2cm}

  \end{titlepage}

   \tableofcontents


\section{Introduction }

\begin{figure}[h]
\begin{center}
 \includegraphics[scale=.4]{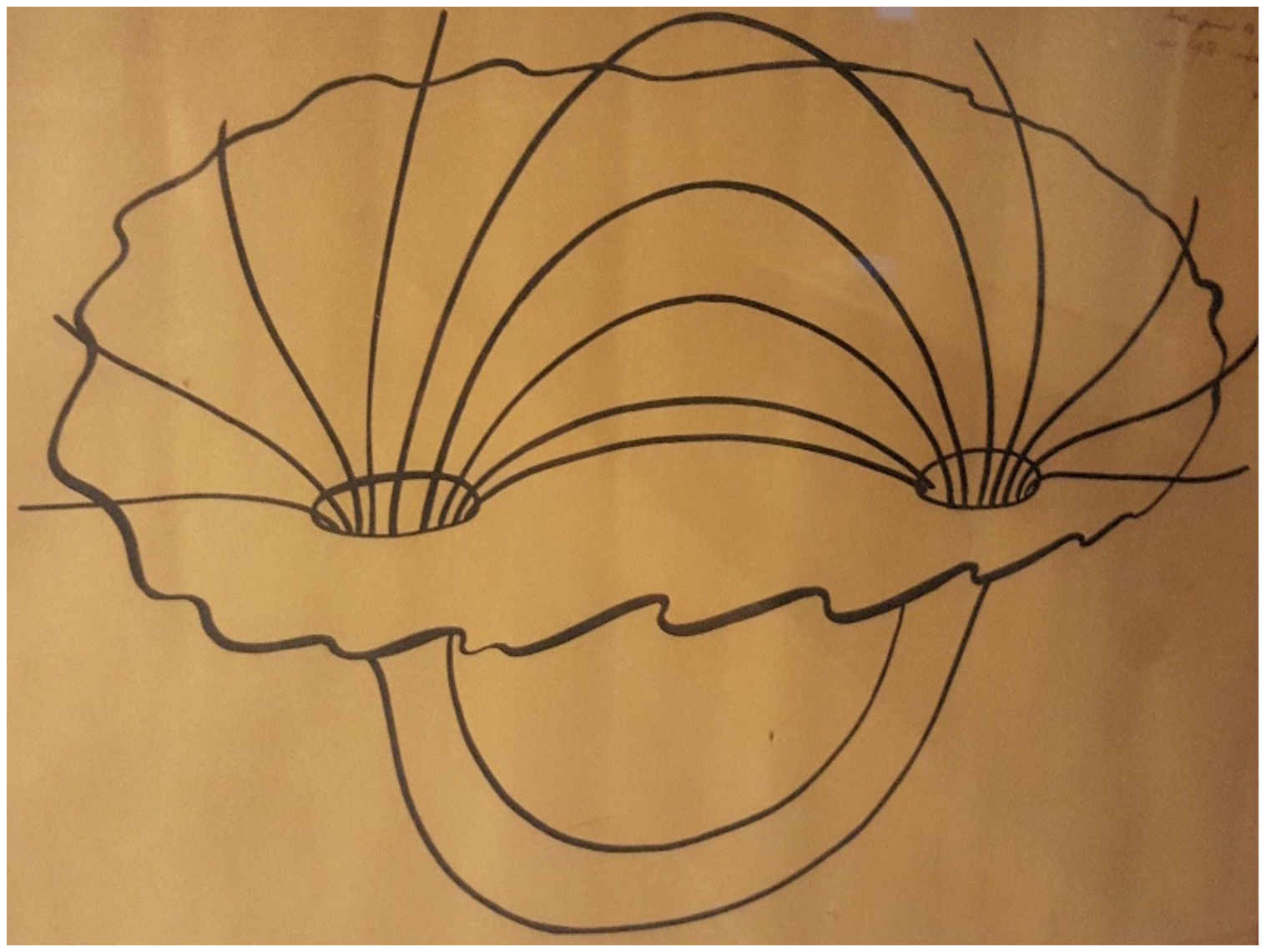}
\caption{Schematic form of the solution we discuss here. It is a traversable wormhole threaded by magnetic fields. This drawing was made by John Wheeler in 1966.  }
\label{Wheeler}
\end{center}
\end{figure}

\begin{figure}[h]
\begin{center}
 \includegraphics[scale=.5]{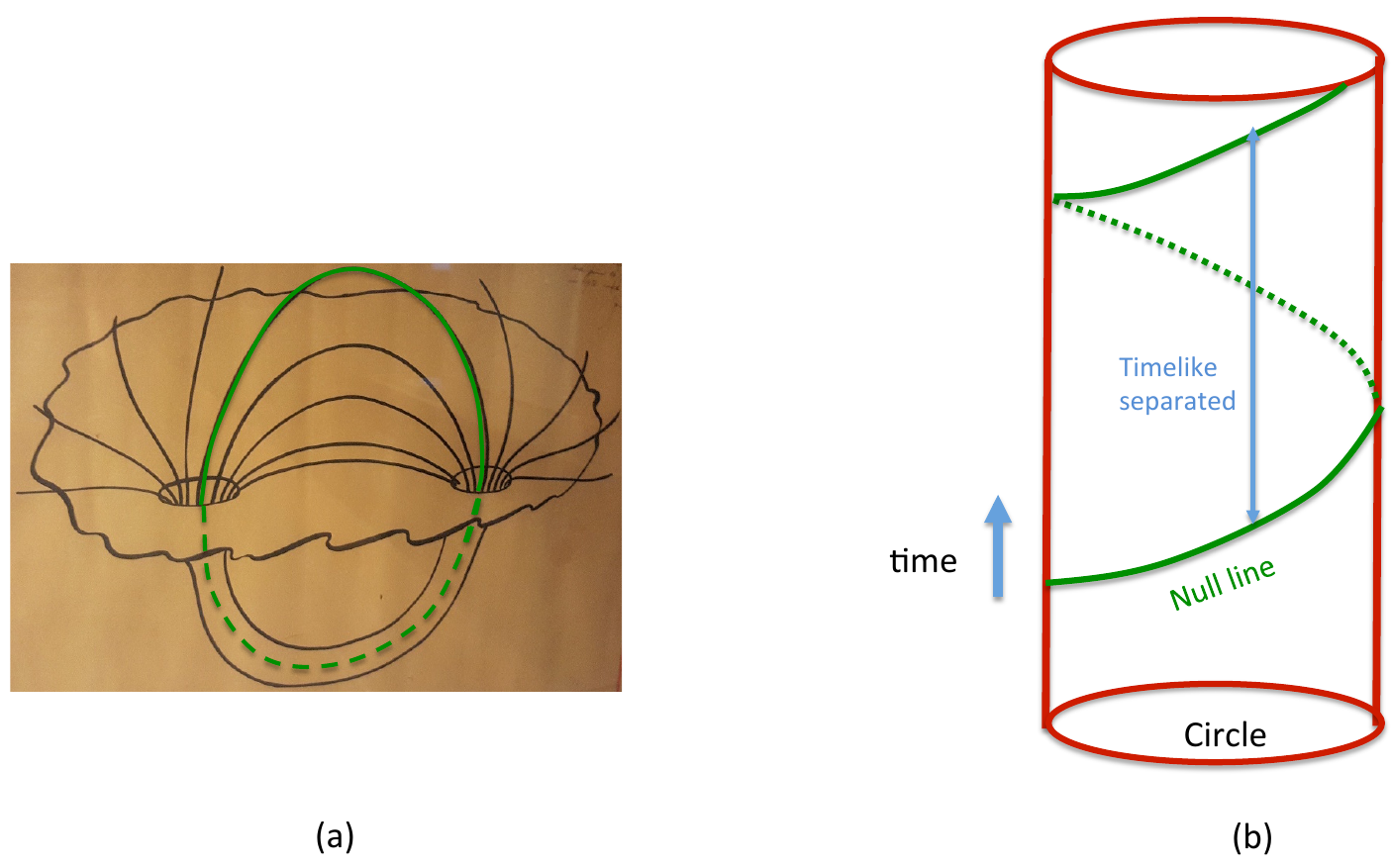}
\caption{  In (a) we see a particular field line highlighted in green. There is a corresponding lowest Landau level state
 localized around this field line. It describes a massless 
two dimensional field moving along the field line, which has the topology of a circle. In (b) we see a cylindrical spacetime. The null line that wraps around the circle is 
{\it not} achronal, since we can see that there are points that are timelike separated along this null line. The average null energy  along this line is negative.     }
\label{Cylinder}
\end{center}
\end{figure}

 It has been a long-standing speculation that wormhole configurations might exist, see
 figure  
 \ref{Wheeler}, for example. 
 Traversable short wormholes that join very distant points in space, and 
 would lead to causality violation, are not allowed by the Einstein equations combined with the achronal average null energy condition \cite{Galloway:1999bp,Galloway:1999br,Gao:2000ga,Graham:2007va}\footnote{ 
 The achronal average null energy condition is expected to be obeyed in general quantum field theories. It has recently been proven in general  flat space theories \cite{Faulkner:2016mzt,Hartman:2016lgu}. Achronal means that no two points along the null line  are  timelike separated. They are, in a sense, the fastest null lines.  Non traversable wormholes  are easily realizable in term of  two sided black hole solutions \cite{Fuller:1962zza}.  }. 
 However, long traversable wormholes are in principle allowed. By ``long'' we mean that it takes longer to go through the wormhole than through the ambient space. 
  Such solutions  should crucially use quantum  effects because they are not allowed in classical physics, which obeys the average null energy condition\footnote{
  The fact that   long wormholes are not allowed  in classical physics  can be understood as follows. In classical physics the equations are local, so we can 
  go the  covering space, where it joins two different universes. In this case, the wormhole becomes short, since the  two different universes are infinitely far, so it
  should be forbidden by causality. We thank A. Wall for this remark. Our solution will use Casimir energy, which depends on the actual topology of space, so we cannot go to the covering space.}. 
   However, it has been difficult to find a controllable example. 
   Here we provide an example by considering a solution that is close to a pair of extremal magnetically 
   charged black holes. A single extremal charged black hole develops an infinitely long throat with the geometry of $AdS_2 \times S^2$. We consider a configuration where  two such throats are joined to each other by a spacetime that again looks like $AdS_2 \times S^2$, but  in global coordinates. The combined configuration has  a time translation symmetry that corresponds to  the global time translation of $AdS_2$ in the throat region. It has no horizon.
    In the presence of the magnetic field,  a massless charged fermion gives rise to a series of Landau levels. The lowest Landau  level has zero energy in the sphere directions and gives rise to massless fields in the 
   radial and time directions. This Landau level has degeneracy $q$, where $q$ is an  integer specifying the quantized magnetic flux through the sphere. A crucial idea in this paper is that,  for large $q$, $q \gg 1 $, 
   a single four dimensional fermion field gives rise to $q$ two dimensional fields. 
   This large number of fields enables parametric control of the solution. 
   Each of the states within the lowest Landau level can be viewed as
    localized along a magnetic field lines.  The corresponding two dimensional field moves on the spatial circle described by this magnetic 
   field line. A two dimensional fermion on a circle gives rise to a negative Casimir-like  vacuum energy. The vacuum stress tensor develops a negative null-null component leading to a violation of the  average null energy condition along the null lines wrapping around the cylinder. 
 Of course,   these null lines are {\it not}  achronal,  see figure \ref{Cylinder}. 
   This negative energy makes it possible to find a solution with a geometry close to 
   global $AdS_2$ in the throat region. 
   The Einstein equations in the throat region determine the length of the throat. 
   In the ambient space, the two mouths of the throat look like a pair of extremal black holes with opposite charges. We can create a long lived solution if we imagine that they are rotating around each other.  By adjusting the distance between them, we can make sure that they rotate slowly enough so that the throat is not de-stabilized. Rotation implies that the configuration emits gravitational and ``electromagnetic'' radiation which will cause the two mouths of the wormhole  to eventually merge. This is a slow process and the wormhole can exist for a relatively long time. The  configuration is rather fragile. If enough  energy is sent through the wormhole mouth, a pair of horizons will develop.  However, the wormhole can be safely explored by sending low energy waves, whose propagation will display the characteristic features of the wormhole configuration. 
   
     The wormhole solution can be viewed as 
   a pair of entangled black holes \cite{Israel:1976ur,Maldacena:2001kr,Maldacena:2013xja}. Similar solutions were found in  \cite{Maldacena:2018lmt} 
   (following \cite{Gao:2016bin,Maldacena:2017axo}) by postulating an interaction 
   between the fields on the two asymptotic regions of $AdS_2$. In this four dimensional configuration this interaction is generated automatically by the exchange of massless fields in the bulk. 
   
    The elements we used are rather general and solutions of this type are expected in any theory that contains these elements. In particular, the Standard Model of particle physics contains these elements.  The $U(1)$ gauge field can be considered to be the weak hypercharge gauge field.  And the overall size of the configuration, namely the distance between the two wormhole mouths, can be taken to be smaller than the electroweak scale to ensure that 
   the fermions can be approximated as massless. 
   
   There are many possible variations of the basic construction, such as putting it in $AdS_4$, etc, which we will not explore in detail here. 
   
   Another very interesting question which we will not address is whether the wormhole can be reached starting from two separate near extremal black holes.  
   
   Interesting examples of traversable wormholes supported by Casimir energy were also 
   recently discussed  in \cite{Fu:2018oaq}. 
   
  This paper is organized as follows. In section two we review the geometry of extremal and near extremal charged black holes. We also consider a charged fermion propagating on this background. In section three we consider a pair of pointlike magnetic charges and we study the geometry of the field lines.
  In section four we discuss the general strategy for finding the solution by splitting it into three different regions.  
  In section five  we discuss the wormhole  region, the Casimir energy, and we show that enforcing Einstein's equations fixes the length of the wormhole. 
    In section six we discuss the rotating configuration of the two mouths. 
    In section 8 we discuss some aspects of this type of solutions in the Standard Model. In section 9 we present some conclusions.

\section{Single Magnetically charged black hole } 

\subsection{The theory} 
 
 In this paper we will consider a theory described by Einstein gravity plus a $U(1)$ gauge field coupled to a set of massless Weyl  fermions. In order to have no anomalies
 we need that  the sum of the cube of the charges and well as the sum of the charges is zero. 
 As a simple example, we can consider a single charged Dirac fermion with charge one. We will discuss this case for most of the paper. The action is 
  \be \la{Lag}
 I =  \int d^4 x \sqrt{g} \left[ { R \over 16 \pi G_N } - { 1 \over 4 g^2 }   F^2 +   i  \bar \chi (\slashed{\nabla } - i \slashed{A} ) \chi \right]
 \ee
 We assume that $g$ is small so that loop corrections are suitably suppressed.

   \subsection{Single charged black holes }

\begin{figure}[h]
\begin{center}
\includegraphics[scale=1.0]{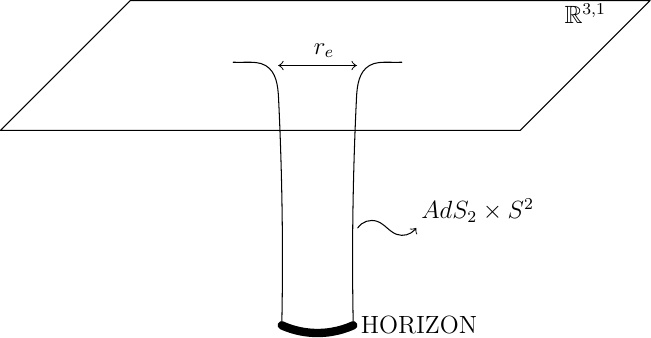}
\caption{ Schematic form of a near extremal black hole geometry. As we approach extremality,  a long ``throat'' with an $AdS_2\times S^2$ geometry develops.
 }
\label{NearExtremal}
\end{center}
\end{figure}

   We start by recalling the form of the magnetically charged black hole solution, with  metric 
   and gauge field given by 
\bea  \la{Metr}
ds^2 &=&  -\left( 1- \frac{2 M G_N}{r} +\frac{r_e^2}{r^2} \right) dt^2 + \left( 1- \frac{2 M G_N}{r} +\frac{r_e^2}{r^2} \right)^{-1} dr^2 + r^2 ( d\theta^2 + \sin^2 \theta d\phi^2 )~~~~~~~~ \\
A &=&  \frac{q}{2} \cos{\theta} d\phi  ~,~~~~~~~~~~~
r_e^2 \equiv { \pi   q^2 l_p^2 \over g^2 }  ~,~~~~~~l_p \equiv \sqrt{G_N} 
\eea
where $M$ is the mass and  $q$ is an integer, giving us the charge of the black hole.  Here $g$ is the coupling constant of the $U(1)$ gauge field. In the quantum theory, $g$ is the running  coupling evaluated at distance scale $r_e/\sqrt{q}$, which is the scale set by the magnetic field.  
The horizon is located at $r=r_+$, where $r_\pm$ are given by:
\be
r_\pm = M G_N \pm \sqrt{M^2 G_N^2-r_e^2}
\ee
The temperature and entropy are given by 
\bea
\label{eq:thermo}
T  =  \frac{r_+-r_-}{4 \pi r_+^2} ~,~~~~~~~~~
S = \frac{\pi r_+^2}{G_N}
\eea
At extremality, $T\to 0$ and  $r_+ =r_- = r_e$. 
Near extremality,  for small  temperatures, we can expand
\be \la{NeEn}
\begin{split}
M &= \frac{r_e}{G_N} + \frac{2\pi^2 r_e^3 T^2}{G_N} + \cdots  \\
S &= \frac{\pi r_e^2}{G_N} + \frac{4 \pi^2 r_e^3 T}{G_N} + \cdots 
\end{split}
\ee
The near horizon geometry of these black holes is approximately given by an $AdS_2 \times S^2$  geometry
\bea \la{RindlerAdS}
ds^2 &=& r_e^2 \left[ - d\tau_r^2 ( \rho_r^2 -1) + { d\rho_r^2 \over {\rho_r^2 -1} }  +  (d\theta^2 + \sin^2 \theta d\phi^2 ) \right]  ~,
\\
& ~& ~~~~~~\tau_r = 2 \pi T t ~,~~~~~~~ \rho_r = { (r -r_e) \over 2 \pi T r_e^2 }  ~,~~~~~{\rm for } ~~~ r-r_e \ll r_e  \la{TempRes}
\eea
where we have also given the relation between the near horizon coordinates and the ones in \nref{Metr}.  We see that this geometry is a good description when 
$ r-r_e   \ll r_e$ and also when $T r_e \ll 1$. This geometry connects with the flat space region via a transition region which is located where $r\sim r_e$. 
See figure \ref{NearExtremal}. 

   \subsection{Charged fermions moving in a magnetic black hole  }

   We now consider a massless charged fermion moving in the magnetically charged black hole background \nref{Metr}. 
   The basic physics is the following. We have a magnetic field on the sphere. This leads to a series of Landau levels. The energy of the Landau levels has an orbital 
   contribution and a magnetic dipole contribution. For fermions, these two precisely cancel for the lowest Landau level. 
   So we get  states with zero energy on the sphere.
   In fact,  this lowest level has degeneracy $q$, which is also related to the angular momentum, $j$,  by $2 j +1 = q$. 
   Here we assumed that the fermion has charge one. 
   Each of these states gives rise to a massless
   two dimensional field in the radial and time directions. 
   In conclusion, a four dimensional chiral fermion gives rise to $q$ massless two dimensional chiral fermions, whose two dimensional chirality 
   depends both on the four dimensional chirality and the sign of the charge\footnote{This plays an important role in the Callan Rubakov effect
   \cite{Rubakov:1981rg,Rubakov:1982fp,Callan:1982au,Callan:1982ac,Callan:1982ah}. Charged fermions in 
   magnetic black holes were discussed previously in \cite{Alford:1992ef}.}. 
   
   The fact that we get zero energy states on the sphere can be understood in terms of anomalies \cite{Ambjorn:1992ca}. Viewing the $U(1)$ gauge field 
   as a fixed background field, 
  the massless  four dimensional fermion has a $U(1)_V \times U(1)_A$ symmetry. 
   The four dimensional anomalies involving $U(1)_A$  imply that we should have two dimensional anomalies. So we should have two dimensional massless chiral fermions after reducing on the sphere. 
   This can be seen by dimensionally reducing the 
   anomaly polynomial of the four dimensional theory to the two dimensional one, which acquires an overall factor of $q$. 
    This is consistent with  a four dimensional Dirac fermion giving rise to $q$
    two dimensional Dirac fermions. 
   
   So far we have ignored the dynamics of the gauge field. It turns out that the dynamics of the gauge field gives a small mass to some of the two dimensional
    the fermions. We will ignore
   this effect here because it is subleading for large $q$.  Further discussion can be found in Appendix \ref{MassGen}. 
   
   Since this discussion can sound a bit abstract, it is also useful to explicitly solve the Dirac equation. 
   Since the massless Dirac equation (and the Maxwell field) are Weyl invariant we can scale out the metric of the two sphere as an overall factor, so that the size of the 
   sphere is constant. This amounts to dividing the metric \nref{Metr} by  $r^2$. We can then factorize the four dimensional spinor as 
   \be
   \chi = \psi \otimes \eta 
   \ee
   where $\eta$ is a spinor on the $S^2$ and $\psi $ a spinor in the other two directions. 
   The lowest Landau level corresponds to a negative  chirality spinor $\eta_-$ that obeys the two dimensional massless Dirac equation on the two sphere with a magnetic field,  $\gamma^\alpha (\nabla_\alpha - i  A_\alpha) \eta =0$. These solutions have the form 
   \be
   \eta_- \propto  e^{ i m \phi}   \left( \sin { \theta \over 2} \right)^{j -m} \left( \cos{ \theta \over 2 } \right)^{j + m } ~,~~~~~ j = \frac{q-1}{2} ~,~~~~~ - j \leq m \leq j 
   \ee
  where $m$ is the $J^3$ quantum number. 
  See appendix C for more details. 
   
 Note that for large $q$ these wavefunctions are well localized near  $\theta_m$ given by 
  \be \la{thetaA}
  \cos\theta_m =  { m \over j}  ~,~~~~~~~~~ (\Delta \theta)^2 \sim { 1 \over q } ~,~~~~~~~~~ q\gg 1
  \ee
  where we also indicated the rough size of the wavefunction around the central value. These wavefunctions are extended along the $\phi$ 
   direction of the sphere.  
  
As we said before,  each of these modes on the sphere gives rise to a two dimensional massless mode $\psi_m$ on the $r$ and $t$ directions. 
 Again, this two dimensional system is conformal invariant, so we can ignore the overall factor in the two dimensional metric. 
 In other words, these fermions propagate on a space with the metric   
   \be \la{RescaL}
   ds^2_2 =  |g_{tt} | \left[ -dt^2 + dx^2 \right] ~,~~~~~{\rm  where} ~~~~ 
   dx = \sqrt{ g_{rr} \over -g_{tt} }dr
  \ee
    
  As a final comment let us note that if we consider a scalar field instead of a fermion field, then the Landau levels all have positive energy, with the lowest one 
  having an energy of order $E \sim \sqrt{B}$. This is   also the order of magnitude of the energy for all the other non-zero energies in the case of the fermion field. 
   In the throat region, this leads to modes which have masses in $AdS$ units of the form $m^2 R_{AdS_2}^2 \propto q$. These relatively large masses lead to fields
   that decay quickly in the $AdS_2$ region and do not give rise to significant Casimir energy. 
       If we had a charged spin one field, then we would have an instability. This arises when the magnetic field is part of a non-abelian gauge group and it represents
       the fact that the magnetic field can be screened. This is why we insisted that the gauge field is $U(1)$.

 \section{Two oppositely charged black holes } 
    
\begin{figure}[h]
\begin{center}
\includegraphics[scale=0.7]{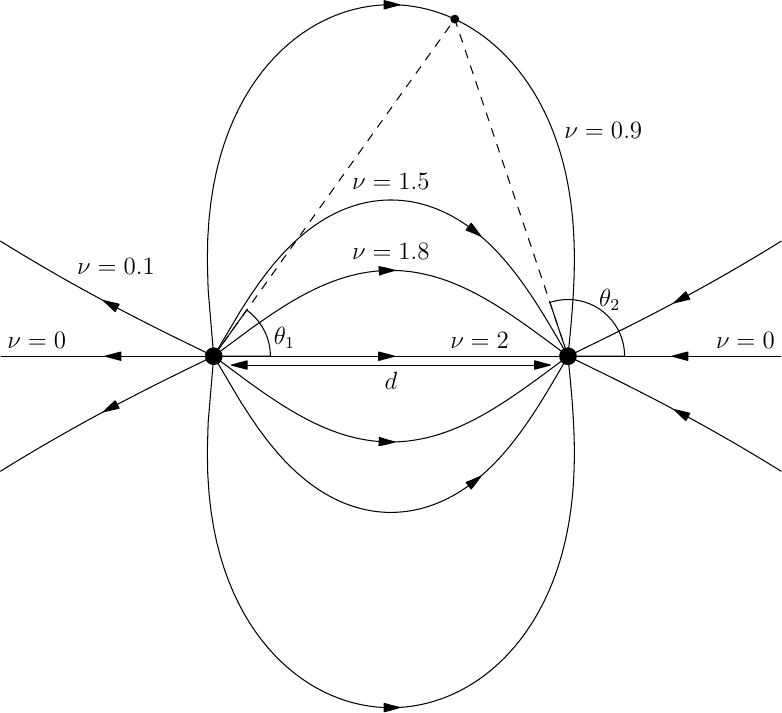}
\caption{ Magnetic field lines for two sources at distance $d$. We have plotted the magnetic field lines for a few values of $\nu$ in equation \nref{MFline}. 
 We have also defined the angles $\theta_1$, $\theta_2$.}
\label{FieldLines}
\end{center}
\end{figure}

  We now consider two oppositely charged magnetic black holes.  Let us imagine that the two black holes are at some relative distance $d$ that is large
  compared to their size, $d \gg r_e$. 
  First we will ignore the motion of the black holes, which will be relatively slow, and focus on the massless fermions, which can move at the speed of light. 
  At distances large compared to $r_e$, we can approximate the black holes as two magnetic monopoles in flat space. These produce a magnetic field simply 
  given by the vector potential (see figure  \ref{FieldLines})
  \be \la{VecPo}
  A = {q \over 2 } ( \cos \theta_1 - \cos \theta_2 ) d\phi 
  \ee
  where $\theta_1 $ and $\theta_2$ are the angles relative to the $\hat z$ axis, which is the axis joining the two black holes and $\phi$ is a azimuthal angle. 
  The configuration is rotational invariant
  around this axis.
  In order to find the form of the magnetic field lines it is convenient to write 
the  spatial metric in polar coordinates,  $ds^2 = d\rho^2 + dz^2 + \rho^2 d\phi^2$.  The magnetic field can then  be written as 
\be
\la{MagFi}
B_\rho = { \partial_z A_\phi  \over \rho} ~,~~~~~~B_z = - { \partial_\rho A_\phi \over \rho }  ~,~~~~~~~~B_\phi =0
\ee
The equations for the field lines should be such that the tangent vector is along $\vec B $. 
We see that if we consider the solutions of the equation $A_\phi =$constant, then its gradient is normal to $\vec B$ due to \nref{MagFi} and therefore
the tangent vectors to the curve are along $\vec B $.  
Then the equations for the field lines are at $\phi=$ constant and 
  \be \la{MFline}
  \cos \theta_1 - \cos \theta_2 = \nu   ~,~~~~~~~  0 \leq \nu \leq 2  ~,~~~~~~\nu = { j+ m \over j } 
  \ee
   where we also indicted the relation between $\nu$ and the value of the angular momentum of a fermion at that angular position near each of the black holes, as 
   computed via \nref{thetaA}.

   \subsection{Fermions outside the black holes }

  We now consider the massless fermions moving in the magnetic field described in the previous section.  
   We are now interested in the solutions at distances comparable to the distance, $d$,  between the
  two sources. We want to focus on solutions with low energies,    low compared to $\sqrt{q}/d$. For this reason we are interested in the lowest Landau levels. 
  Near each of the sources the solutions are approximately the ones described above, with the fermion wavefunctions localized on the field lines. 
  In fact, we can make the approximation that the fermions continue to be localized along the field lines throughout the whole intermediate region. This amounts 
  to solving the Dirac equation in a WKB-type approximation.  
  Quantitatively, the condition that massless fermions follow the magnetic lines can be stated as follows. First of all, for massless fermions the wave function
will be localized over distances $\sim 1/\sqrt{B}$. We demand that the magnetic field changes adiabatically over this distance:
\be
\frac{1}{\sqrt{B}} { 1\over B} \frac{\partial B}{\partial r} \ll 1
\ee
Since the magnetic field goes as the dipole field, namely, as $\sim qd/r^3$, this implies that the fermions will follow the lines up until distance $r \sim qd$.
Using the result of Appendix \ref{flines} we see that this distance corresponds to $\nu$ near 0: $\nu \sim 1/q$. We conclude that only a small fraction of fermions
will not follow the magnetic lines and fly away to infinity. 
See appendix \ref{potential} for further discussion. 

  The fermion with angular momentum $J_z=m$ follows a magnetic field line obeying \nref{MFline}. The length of this magnetic field line is 
    is proportional to $d$ times a function of $\nu$ 
   \be \la{LengthF}
   L_{\rm flat} = d \,  f(\nu) ~,~~~~~  
   \ee
where precise form of $f(\nu)$ is given in appendix \ref{flines}, eqn. \nref{Deffnu}. 
   This is also the length in the $x$ coordinate defined in \nref{RescaL} because $g_{tt}=1$ in this flat space region we are considering.

\section{General strategy} 

\begin{figure}[h]
\begin{center}
\includegraphics[scale=0.5]{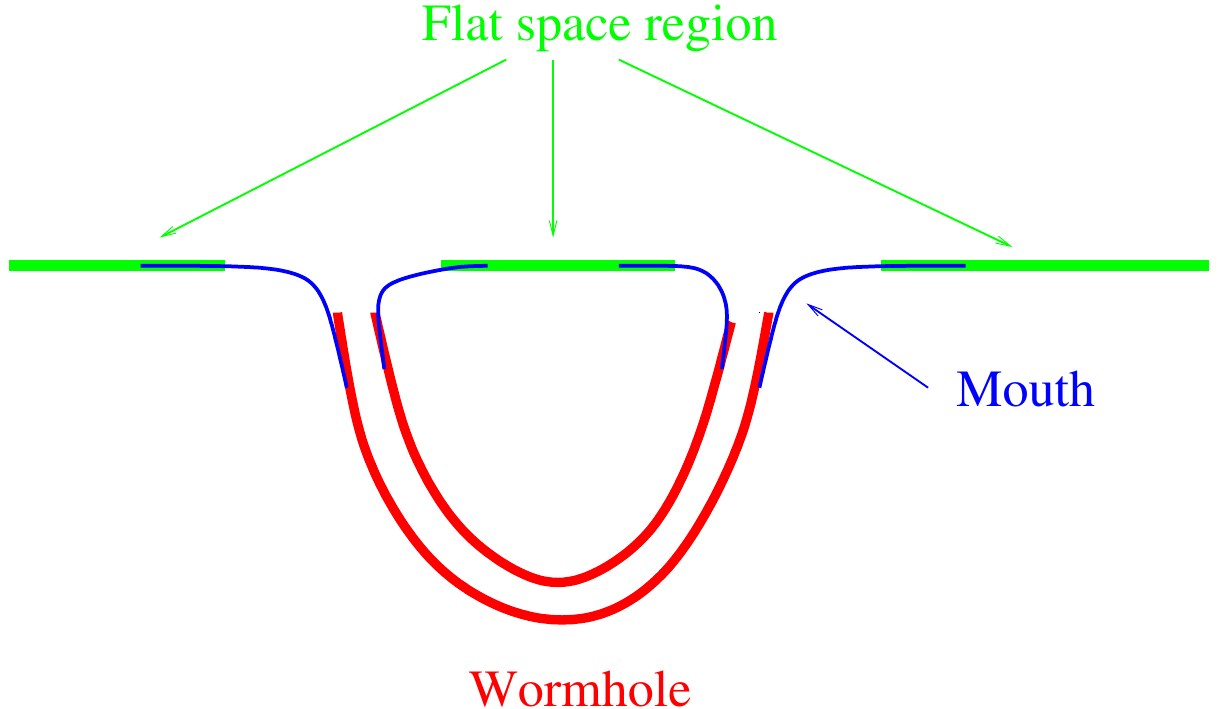}
\caption{ We separate the solution into three overlapping regions: 1) the wormhole, 2) the mouth and 3) the nearly flat space region. Within each region we describe
the solution using a different metric which   coincide with that of the next region in the overlapping regions of validity. 
 }
\label{Regions}
\end{center}
\end{figure}

Our  strategy for finding and describing the solution is the following. 
We   cut the solution in three pieces: 1) the wormhole, 2) the mouth, and 3) the flat space region, see figure \ref{Regions}. We approximate the solution in different ways in each 
region. These approximations have overlapping regions of validity so that we can patch
the solutions. In the wormhole region we   approximate the solution as $AdS_2 \times S^2$ plus a small deformation. In region 2) we approximate it as the mouth 
of the single black hole solution \nref{Metr}.  
 It   turns out to be a black hole with a mass slightly  below extremality. 
 In 3), the flat space region,  we approximate the two mouths as two
  point-like masses and charges, incorporating   their  almost Newtonian dynamics, emission of radiation, etc.

\section{The wormhole region} 
 
In this section we find the solution in  the wormhole region, see figure \nref{Regions}. 
We  do this in a few steps. First we  write down the geometry of the wormhole as $AdS_2 \times S^2$, which is a solution of the leading order
Einstein equations with gravity and the magnetic field. It is not quite a solution of those equations once we  deform it so that 
 each of the two boundaries turns into mouths connecting the throat into flat space. After this deformation   
 we have one free parameter,  called $\ell$, which parametrizes the length of the wormhole. 
We  then consider a massless fermion in this geometry and argue that  it generates negative energy. This negative energy then produces a solution with 
the length $\ell$ fixed  to a special value.  First we     show this from a quick  energy minimization argument. Then we analyze in detail 
Einstein's equations in the wormhole  region.  
     
  \subsection{The geometry} 
  
\begin{figure}[h!]
\begin{center}
\includegraphics[scale=.6]{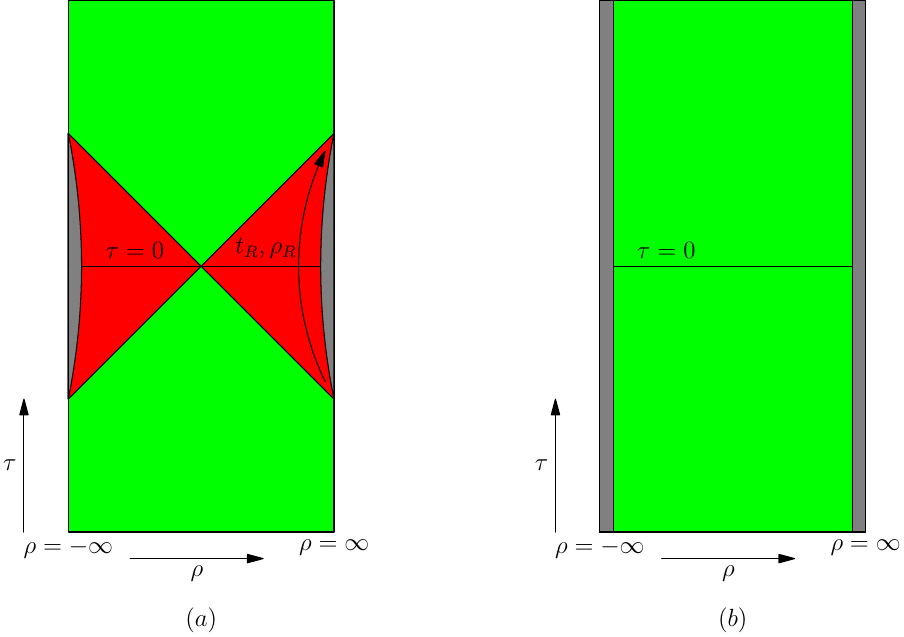}
\caption{ In (a) we display  in red the region corresponding to the near horizon region of a near extremal black hole. The two triangles correspond to the two exteriors of the maximally extended black hole. The regions shaded in grey should be cut out, and instead replaced by 
a geometry that connects to flat space. These red regions correspond to a Rindler-like portion of global $AdS_2$ space. In (b) we consider the full $AdS_2$ space in global coordinates. We cut out the shaded regions near the
 boundaries. We will then patch that geometry to flat space. We have also highlighted a $\tau =0$ slice which is very similar for the geometries in (a) and (b).     }
\label{Coordinates}
\end{center}
\end{figure}

   We will now make a precise ansatz for the wormhole geometry that joins the two throat regions. As a first approximation, the
    geometry is that of $AdS_2 \times S^2$ 
   \be
   ds^2 = r_e^2 \left[ - (\rho^2 +1) d\tau^2 + { d \rho^2 \over (\rho^2 +1) } + ( d\theta^2 + \sin^2 \theta d\phi^2 ) \right] \la{GlobalAdS}
   \ee
   This is a solution of Einstein's equations with a magnetic field on $S^2$. We know this since it is a limit of the original charged black hole solution 
   \nref{RindlerAdS}. 
   \nref{GlobalAdS}  is a different coordinate system for $AdS_2$ compared to \nref{RindlerAdS}. 
   \nref{GlobalAdS}  is the so called ``global'' coordinate system which covers the whole Penrose 
   diagram of $AdS_2$, see figure \ref{Coordinates}. We want to match each of the two asymptotic regions to each of the two black holes, see figure 
   (\ref{Coordinates}b). 
   So for large $\rho$ we match the geometry \nref{GlobalAdS}  to the small $r$ region of the extremal black hole geometry, \nref{Metr},  via 
   \be
    \tau ={  t \over \ell } ~,~~~~~~~~ 
    \rho = { \ell ( r-r_e ) \over r_e^2 }  ~,~~~~~~ {\rm for}~~ 1 \ll \rho  ~,~~~~~~{ r- r_e \over r_e} \ll 1 ,~~~~~1 \ll { \ell \over r_e }  \la{Resca}
    \ee
    where $\ell $ is a free parameter that we will soon fix. 
    Since the throat opens up at $r-r_e \sim r_e$ we see that we can trust the $AdS_2$ geometry until 
     $\rho_{\rm cutoff} \propto  { \ell \over r_e} $. We assume that 
    $\ell/r_e\gg 1$ and we will check it later. 
    Here $\ell$ is a length like quantity that plays a role similar to the inverse temperature $\beta = 1/T$ in 
    \nref{TempRes}, in the sense that it provides a rescaling between the $AdS_2$ time and the asymptotic time. 
    Notice that the time direction never shrinks  in \nref{GlobalAdS}, so that this configuration is {\it not} a 
    black hole, since it does not have a horizon. And this spacetime is at zero temperature (for now).

 It is interesting to calculate the rescaled length of the wormhole in the $x$ coordinates introduced in \nref{RescaL}. We find 
 \be \la{Length}
 L_{\rm throat}  =  \Delta x \sim  \int_{-\rho_{\rm cutoff} }^{\rho_{\rm cutoff} }  d\rho { \ell \over  ( 1 + \rho^2 ) }  \sim  { \pi \ell }
 \ee
 Therefore we can interpret $\ell$ as setting  the ``length'' of the wormhole. Notice that this is {\it not} the proper length. It is the effective length that 
 a two dimensional massless field will effectively see, after we normalize the time coordinate as in the asymptotically flat region. 
 Note that \nref{Length} is also the time that it takes to go through the wormhole, as measured in terms of the asymptotically flat spacetime coordinate $t$. 
In other words, if an outside observer sends a signal into one side at $t=0$, he or she will see it emerge from the other side at time $t = \pi \ell$. 
 $\ell$ also sets the energy gap for excitations deep inside the wormhole. For example, we can  consider a massless four dimensional field,
  which we expand in Kaluza Klein modes 
 on the $S^2$. Then,   these excitations have energies of order one measured in terms of the time $\tau$. However, 
 they have energies of order $1/\ell$ measured in terms of the outside time $t$. So we can say that the energy gap for the wormhole geometry  is 
 \be
 E_{\rm gap} \propto {1 \over \ell }  \la{ThGap}
 \ee

 \subsection{Negative energy } 
 
  Solutions with the shape that we have discussed so far cannot exist without negative energy. However, we have set up the whole configuration so that 
  we have a sizable source of negative energy. This comes from the effectively massless two dimensional fermions. These are modes that move along the magnetic 
  field lines. They come out of one the throats, moving along the field lines in the flat space region, they go into the other throat and then through the wormhole 
  in \nref{GlobalAdS} they emerge again in the original throat. Therefore each massless mode lives on a circle, parametrized by coordinates $t$ and $x$, 
  \nref{RescaL},  of total length 
  \be
   L \equiv \oint dx \sim  L_{\rm throat} + L_{\rm flat}(\nu) ~,~~~~~L_{\rm throat} = \pi \ell ~,~~~~~L_{\rm flat} = d \, f(\nu) 
  \ee
  where $\pi \ell$ is the contribution in the connecting wormhole region and $L_{\rm flat}(\nu)$ is the distance \nref{LengthF}. 
  First we will assume that 
  \be
  \ell \gg d  ~,~~~~~~   \la{LaDi}
  \ee
  This implies that for most of the field lines we have that $\ell \gg L_{\rm flat}(\nu)$ and  
  $L = \pi \ell $.  After we find the value of $\ell$, we will check this assumption, \nref{LaDi}. 
  
  If we had $q$ complex fermions with both left and right moving components moving on a circle of length $L$  we would have a ground
  state energy 
  \be \la{CasFlat}
  E_{\rm flat~cylinder}  = - { q \over 12 } { 2 \pi \over L }  =   - { 1 \over 6 } { q \over \ell } 
  \ee
  
  It is tempting to say that this is also the energy in our case.  
  However, in the wormhole region, the spacetime where these fermions live is not flat, it is only conformally flat. 
   So we  need to take into account the two dimensional conformal anomaly. 
  One quick way to derive the effect of the anomaly is to note that for the case of $AdS_2$ with reflecting boundary conditions at both  boundaries we 
  do not expect to see negative energy. In this case the field configuration and the boundary conditions are  $SL(2)$ invariant so we cannot have a null-null component
  in the stress tensor. On the other hand, on a flat strip of length $L$ we have a negative energy of the form 
  \be
  E_{\rm flat~strip} =- {q \over 24 } { \pi \over L } 
  \ee
  This means that the conformal anomaly contribution must be such that it cancels this precisely, see appendix \ref{ConfAn}.
  Therefore, in our case, we have a total negative energy of the form 
  \be \la{CasW}
  E_{\rm wormhole}  = + { q \over 24 } { \pi \over L } -  { q \over 12 } { 2 \pi \over L }  = - { q \over 8 \ell } 
  \ee
  
  This is the total energy, but it could be also viewed as generated by a   stress tensor of the form 
  \be  \la{Stre}
  \hat T_{tt} = \hat T_{xx}   =  - { q \over 8 \pi \ell^2} { 1 \over 4 \pi r_e^2 } ~,~~~~~~~\hat T_{\alpha \beta } \equiv T_{\alpha \beta} - \half g_{\alpha \beta} T_{\alpha}^{\, \alpha} ~,~~~~~~\alpha, \, \beta = t, \, x
  \ee
  where we have extracted the two dimensional trace contribution,  which also is related to the conformal anomaly.   This trace part can also have contributions from the
  other four dimensional modes. 
  
 When we wrote \nref{CasFlat} we had implicitly assumed that the fermions have anti-periodic boundary conditions around the cylinder. In fact, this 
 corresponds to a choice of spin structure. It is part of the specification of the geometry in a theory with fermions. We chose the fermions to be antiperiodic. In this particular case of a single charged fermion, the two choices are continuously connected by another continuous variable which the Wilson line of the 
 gauge field around the circle. Since the Casimir energy depends on this Wilson line, the Wilson line is fixed to the value that minimizes it, which is again the 
 antiperiodic boundary condition. This is related to the fact that the two dimensional gauge field becomes massive in the throat, see appendix \nref{MassGen}. 
  
  One might question whether it is reasonable to ignore all the other quantum effects due to all the other modes of the fermion field, as well as the other modes of
  the gauge field and gravity. The logic for ignoring them is the following. All other modes of the fields are massive in the throat region, with a mass at least
  of order one in units of the radius of the sphere or $AdS$. In the case that we have exactly $AdS_2 \times S^2$,  all these other modes give a stress tensor which 
  respects the $SL(2) \times SU(2)$ symmetries and is 
   proportional to the metrics of $AdS_2$ and/or $S^2$. Such contributions to the quantum stress tensor   slightly change  the radius of the
  $S^2$ or $AdS_2$. This is not a   qualitative change. Therefore we will ignore components of the stress tensor that are proportional to the $AdS_2$ and/or $S^2$ metrics.  When we consider the wormhole geometry, we get nontrivial 
  boundary conditions   that relate  the fields at the two opposite ends of $AdS_2$. This arises 
  because the two mouths of the wormhole are close in the ambient space. However, for massive fields, the effect of this change of boundary conditions dies relatively   
  quickly as we go deeper into the throat. This means that we do not get a significant contribution to the qualitatively important components of the 
  stress tensor from these modes.

 \subsection{Stabilizing the wormhole} 
 
 In this subsection we will determine the length $\ell$ of the wormhole. We will first present a quick argument and then present a longer, conceptually clearer, 
  argument, by solving explicitly Einstein's equations with the quantum stress tensor source. 
 
 \subsubsection{Quick variational ansatz argument } 
 
Notice that the full extended Penrose diagram of a near extremal black hole also contains a   wormhole. As opposed to the one we are considering here, this is 
not a traversable wormhole. However,  if we focus purely on the $t=0$ slice of the geometry we find a configuration similar to the $t=0$ slice of the above 
wormhole if we connect the temperature to $\ell$ with the equation 
\be \la{elT}
 T = { 1 \over 2 \pi  \ell } 
 \ee
 obtained by comparing the rescaling of both time directions. 
 In order to make the connection between the two geometries it is useful to note the explicit change of coordinates between $\tau_r,\rho_r$ in \nref{RindlerAdS}
 and \nref{GlobalAdS}. This is most simply obtained by   expressing the embedding coordinates $X^M$, $X^2 =-1$, in both sets of coordinates.  
 In particular, this implies that 
 \be 
 X^1 = \rho = \sqrt{\rho_r^2 -1} \cosh \tau_r  
 \ee
 As $\rho$ goes from being positive to negative we move through the Rindler horizon, 
   ($\tau_r \to \tau_r + i \pi $). 

We can now set up a variational ansatz for the geometry  as follows. First we ignore the quantum corrections and compute the classical energy by focusing on the 
$\tau=0$ slice and equating the geometry to that of the $\tau_r=0$ of the standard near extremal wormhole. This last energy is simply the energy of {\it two }
near extremal black holes \nref{NeEn} at the temperature \nref{elT}. Quantum effects then add the Casimir energy 
 \nref{CasW} so that we get a total energy (relative to that of two extremal black holes) 
\be
E = { r_e^3 \over G_N \ell^2 } - { q \over 8 \ell } 
\ee
We now set the value of $\ell$ by minimizing this, and we get 
\be \la{Vall}
\ell = 16 { r_e^3 \over G_N q } = { 16 \pi^{3/2}   q^2 l_p\over g^3 } ~,~~~~~~~~~ E_{\rm min} = -  {  G_N q^2 \over 256 \,  r_e^3  } = - { g^3 \over 
256 \, \pi^{3/2}   q l_p  } 
\ee
where we have also written the net result for the binding energy coming from the throat, relative the the energy of two disconnected extremal black holes. 
Since this is the minimum energy configuration, we then expect it to lead to a time independent solution, which is the wormhole we are after. 
 
 Though this procedure seems to be intuitively reasonable, we have not derived it directly from Einstein's equations. Therefore in the next section we 
 rederive this result by directly solving Einstein's equations. 
  
  \subsubsection{Directly solving the Einstein equations with quantum matter} 
  \la{SolvEE}
  
 Here we directly solve the semiclassical Einstein equations. We will divide the problem into two regions. 
  First we consider the classical solution in a region that goes from flat space until deep in the throat. There we will mostly ignore the quantum effects. 
  We make a spherically symmetric ansatz for the geometry 
  \be \la{MetA}
  ds^2 = - A(r) dt^2 + B(r) dr^2 + r^2 d\Omega^2  ~,~~~~~~~~~~~~F =dA = -{ q \over 2} \sin \theta d\theta d\phi
  \ee
  The most general solution that is asymptotically flat has only one integration constant which is the mass parameter in \nref{Metr}. 
  This integration constant will be set after we solve the equations in the wormhole region. 
  \be \la{ABva}
  A = { 1 \over B} = \lll 1 - { r_e \over r } \rrr^2 - { 2 \varepsilon \over r } ~,~~~~~~~~ \varepsilon \equiv  G_N   M - r_e
  \ee
  where $\varepsilon $ is proportional to the difference between the mass and the extremal mass.

 In the wormhole region, we could dimensionally reduce the equations to two dimensions, where we recover the Jackiw-Teitelboim theory
 \cite{Jackiw:1984je,Teitelboim:1983ux,Almheiri:2014cka,Nayak:2018qej}. Then the solution 
 of the equations can be obtained from appendix C of  \cite{Maldacena:2018lmt}. 
  For the benefit of the readers who are not familiar with all that, we will solve them directly here. 
  
 In the wormhole region we can expand the metric as a perturbation away from $AdS_2\times S^2$ 
  \be \la{FdAn}
  ds^2 =  r_e^2\left[ -   (1 + \rho^2 + \gamma )   d\tau^2 +    { d \rho^2 \over 1 + \rho^2 + \gamma } +    (1 + \phi ) ( d\theta^2 + \sin^2 \theta d\phi^2 ) 
  \right]
  \ee
   where $\phi$ and $\gamma$ are small quantities, so that we expand the denominator to first order in $\gamma$. 
The four dimensional stress tensor contains a classical piece from the magnetic field. The quantum part of the stress tensor was computed in \nref{Stre}. 
 
We can now   write the $\rho \rho $ component of Einstein's equation to obtain 
\bea
& ~& R_{\rho \rho } - { 1 \over 2 } g_{\rho \rho  } R - (8 \pi G_N) [ T^{\rm mag}_{\rho \rho  }  + \hat T_{\rho \rho }  ] =0  \rightarrow
 \\
 &~& {\rho \phi'   } - { \phi  } = (8 \pi G_N) (1 + \rho^2)  \hat T_{\rho \rho  } = - {   \alpha \over (1 + \rho^2 ) }  ~,~~~~~~ \alpha \equiv   { q \over 8 \pi } { ( 8 \pi G_N) \over 4 \pi r_e^2 }  \la{MidEq}
 \\
 \rightarrow &~& \phi = \alpha ( 1 + \rho \arctan \rho )    \la{FinSo}
 \eea
 where we used that the quantum contribution to the stress tensor is  
 \be
 \hat T_{\rho \rho}   = { \hat T_{\tau \tau } \over (1 + \rho^2 )^2 }  ~,~~~~~~~ \hat T_{\tau \tau } = \ell^2 T_{tt} =  - { q \over 8 \pi   } { 1 \over 4 \pi r_e^2 } 
 \ee
  where we used  \nref{Stre} and \nref{Resca} to relate the $t$ and $\tau$ variables.  There $\hat T_{\alpha \beta} $ denotes the traceless components in the two dimensional sense. 
  The parts of the stress tensor that are proportional to the $AdS_2$ metric or the $S^2$ metric simply rescale the overall sizes of the $S^2$ or $AdS_2$ and are 
  not interesting for our purposes.

 The most general solution of  \nref{MidEq}  contains also a term linear in $\rho$ which we have set  
  to zero by demanding a $Z_2$ symmetry around $\rho =0$ to obtain \nref{FinSo}. 
  The $\tau\tau$ component of Einstein's equations is then automatically obeyed. In addition, we can look at the sphere components of Einstein's equations. 
  These determine the form of $\gamma$.  Fortunately  we will not need   $\gamma$.

 We now should match the throat solution to the asymptotic solution. We perform this matching at large $\rho$ where 
 \be \la{rvsrho}
 r^2 \sim r_e^2 ( 1+ \phi) ~~\longrightarrow~~~ 2{r-r_e \over r_e } = \phi \sim  { \pi \alpha \over 2 } \rho     
 \ee
 where the first equality is obtained by comparing the sphere part of the metric between   \nref{MetA} and \nref{FdAn}. The second equality comes from
 expanding \nref{FinSo} at large $\rho$. 
  We can now compare the leading order  time component of the 
 metrics to  determine $\ell$
 \be \la{DetL}
  { ( r - r_e) \over r_e^2 } dt = \rho { d \tau}    \longrightarrow   \ell = { dt \over d\tau } = {r_e^2  \rho \over (r-r_e)  } = { 4 r_e  \over \pi \alpha     } = 
  16 { r_e^3 \over q  G_N }  
  \ee
  which agrees with \nref{Vall}. 
  We can also compare the leading expansion of the metric \nref{MetA} \nref{ABva} with \nref{FdAn} (with $\gamma =0$) to obtain 
  \be \la{gttCom}
  r_e^2 (1 + \rho^2) d\tau^2 = \left[{ ( r -r_e)^2 \over r_e^2 } - 2 { \epsilon \over r_e} \right]dt^2 
  \ee
  Using \nref{DetL} and \nref{rvsrho} we see that the $\rho^2$ term is identical to the $(r-r_e)^2$ in the right hand side. Equating the   constant terms on
  both sides we get 
  \be \la{EpsVa}
  E_{\rm min} = { 2 \epsilon \over G_N } = - { 1\over 256 } { q^2 G_N  \over r_e^3 } 
  \ee
  which is the wormhole energy reproducing what we found in  \nref{Vall}. The factor of two comes from the mass correction to each of the two black holes. 
  We have also checked that the radial components of the metrics match. 
 We see that we only needed to use the leading form of the time component of the metric, so that we justify neglecting $\gamma$ in \nref{FdAn}. 
 If you are curious, you can find the form of $\gamma$ in  appendix \ref{gamma}.  
  
  Note that $\epsilon$ in \nref{EpsVa} is smaller than zero, which is less than the extremality bound. For a single black hole this would lead to a naked singularity. 
  However, in our case, with the additional Casimir energy contribution, we find that the geometry of the throat is completely smooth. 
   
 \subsection{Stability of the throat under small perturbations}
  
  It is important to check that the throat is not so long that quantum gravity fluctuations become large. For a near extremal black hole this happens when 
  the temperature is so low that the near extremal entropy is of order one, $r_e^3 T/G_N \sim 1$. Since $1/T$ is setting the length of the throat, and is analogous to $\ell$,
  we expect that quantum fluctuations will become important    
   when $r_e^3 /( G_N \ell) \sim 1$\footnote{ This combination is the effective coupling of the Schwarzian theory that governs the gravitational effects in 
  Nearly-$AdS_2$  \cite{Jensen:2016pah,Maldacena:2016upp,Engelsoy:2016xyb}. }.
   So the solution presented here is valid only when 
  \be \la{QuanSure}
  \ell \ll  q^3 l_p
  \ee
  In \nref{Vall}   we got $\ell \propto q^2 l_p $, which obeys this bound\footnote{When we give the scalings for large charge, using $\propto$, 
   we imagine that $g$ is small and
   fixed as $q $ becomes large. One can also consider theories where $g$ is parametrically small, but we will not discuss it explicitly here.}.

   As we mentioned above,  the throat energy gap is  \nref{ThGap}, $E_{gap} \sim 1/( q^2 l_p)$. 
   However, if we send too much energy into the throat we run the risk of making a near extremal black hole. 
   This happens when we send an energy that is larger than the binding energy of the throat \nref{Vall} which is of order $1/( q l_p)$. 
   While this is larger than the gap by a factor of $q$, it is still a relatively small energy. In fact, an object that has a Compton wavelength of order $r_e$ outside the
   throat will already have this energy!. This means that these wormholes are not safe for human travelers, who will need $r_e$ much larger than their Compton 
   wavelength!. On the other hand, the presence of the wormholes can be detected by sending very low energy waves.  These could be the charged fermion fields
   themselves, for example. They can still make into the throat region  and can come back out the other black hole after a time of order $\ell$, revealing the presence of an energy gap. 
   We can say that the wormhole is traversable for waves, but not for very small and energetic objects. It is a fragile and delicate wormhole. 
   
   \subsection{Wormholes at larger distances, $d > q^2 l_p$ } 
   \la{LargeDis} 
   
  Since we found that $\ell \sim q^2 l_p $ and we had assumed that $d \ll \ell$ \nref{LaDi}, we conclude that the discussion in the preceding sections holds only in the case
  $d\ll q^2l_p$. We can wonder what happens when $d$ is larger than this value. To analyze this we need to take into account the length of the field lines outside the 
  wormhole \nref{Length}. So the total length is $L = \pi \ell + d f(\nu)$, 
   which depends on the particular angular momentum of the charged fermion wavefunction. 
  The Casimir energy will then have a contribution involving $1/L$ which needs to be integrated over $\nu$. This approximation breaks down for some of the fermion wavefunctions that can escape to infinity, but these give a subleading contribution anyway, so we ignore this issue. Notice that when we talk about ``length'' we are 
  referring to length in the coordinate $x$ defined in \nref{RescaL}. 
  We should also take into account the effects of the conformal anomaly inside the throat. This is a positive energy contribution that only depends on the geometry 
  of the throat (and $\ell$), but is independent of the total length of the magnetic field lines. 
  Putting these two contributions together, and adding the classical energy   we get 
  \be \la{EnSu}
  E =  {r_e^3  \over \ell^2 G_N } + { q \over 24\ell } -  { q \pi \over 6 } \int_0^2 { d\nu \over 2} { 1 \over ( \ell \pi + d f(\nu) ) }
  \ee
  The first term is  the classical energy, the second term arises from the conformal anomaly in the throat region,
    and the final contribution is the integral over all the field lines of the Casimir energy.
   
     We can get the value of $\ell $ by minimizing \nref{EnSu}.  Here it will also give the same answer as directly solving Einstein's equations, because the 
    stress tensor is proportional to the $\ell$ derivative of the last two terms in \nref{EnSu}, and, in equation \nref{MidEq}, all that appears is the total integral of the stress tensor over the throat region,  which  ends  up equated to the same $\ell$ derivative of the first term of \nref{EnSu} as it happened in   \nref{DetL}  in section \nref{SolvEE}.

\begin{figure}[h]
\begin{center}
\includegraphics[scale=0.6]{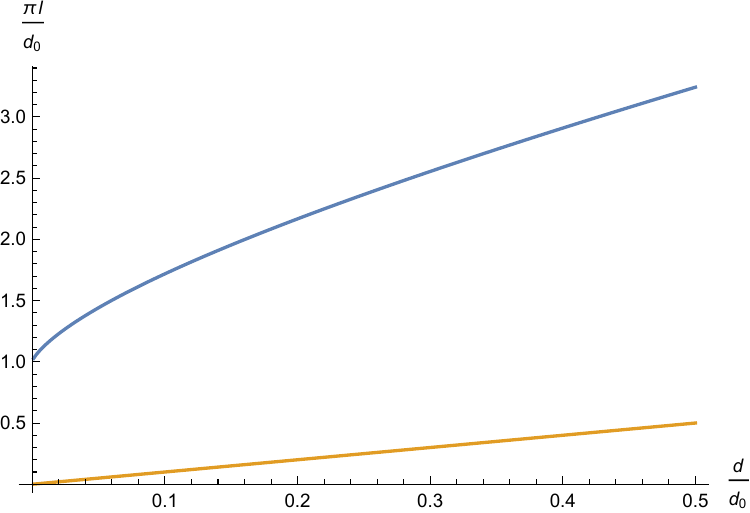} ~~~~
\includegraphics[scale=0.6]{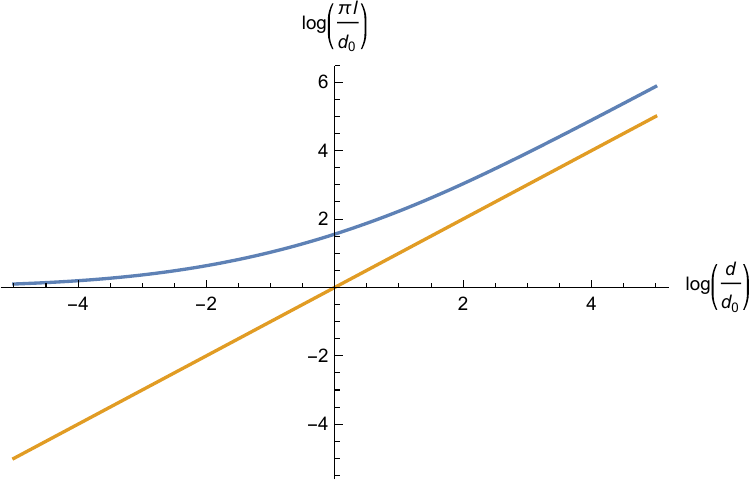} 
\\
~~~(a)~~~~~~~~~~~~~~~~~~~~~~~~~~~~~~~~~~~~~~~~~~~~~~~~~~~~~~~~~(b) 
\caption{ (a) Linear plot of the relation between $\pi \ell$ and the distance $d$. For distances $d\ll d_0$ we see that $\pi \ell/d_0  $ goes to one.  
In orange we see  the line $\pi \ell = d$, which is the causality bound, anything above this line is consistent with causality. 
(b) The same plot in log-log form. For large $d$ we approach a linear relation $\pi \ell \propto d$, with a proportionality constant bigger than one. 
The orange curve is again the causality bound.   }
\label{PlotLvsd}
\end{center}
\end{figure}

    As a toy model, just to get oriented, it is convenient to replace the integral  in \nref{EnSu} 
    by $1/( \pi \ell + d)$ which amounts to approximating the integrand as a constant given by its largest term. 
    Namely it amounts to saying that all fermions go from one mouth to the other by the shortest path, instead of following the field lines. 
    In this case we can minimize the energy by imposing 
    \be
    { \partial E \over \partial \ell } = 0 ~,~~~~\longrightarrow ~~~~
      2 { r_e^3 \over G_N \ell^3 } =    { q \over 6 } \left[  - { 1 \over  4\ell^2 } + { 1\over ( \ell + d/\pi)^2 } \right]  \la{Econ}
      \ee
      In the approximation that we neglect the left hand side we find that 
      \be
      \pi \ell = d 
      \ee
      which is saying that the length through the wormhole is equal to the length through the ambient space. Notice that this result depends crucially on the 
      relative coefficient of  the flat space Casimir energy contribution (the last term in \nref{Econ}) and the conformal anomaly contribution
      (first term on the right in \nref{Econ}). 
      For $d \gg q^2 l_p $ we can safely neglect the left 
      hand side relative to the right hand side of \nref{Econ}. 
      If we included the left hand side of \nref{Econ},  we would get a   bigger distance through the wormhole. 
    We see that including the correct Casmir energy, as in \nref{EnSu}, we are having distances along the field lines  are all greater or equal to $d$, so that the 
    value of $\ell$ that minimizes \nref{EnSu} will be such that the distance through the wormhole is larger than $d$. 
    To study \nref{EnSu} it is convenient to rewrite it as    
  \be \la{Enapp}
  E = { \pi q \over 6 \, d} \left[ 
  { 3 \over 8 } { d_0 \over d \,  y^2 } + { 1 \over 4 y } -    \int_0^2 { d\nu \over 2} { 1 \over ( y  +   f(\nu) ) }  \right] ~,~~~~~~
  y \equiv { \pi \ell \over d } ~,~~~~~~d_0 \equiv { 16  \pi r_e^3 \over q G_N} \propto q^2 l_p
  \ee
   where we defined $d_0$ to be the distance through the wormhole when $d  \ll q^2 l_p $, see \nref{Length},\nref{Vall}. 
  We can now find the solution by minimizing over $y$, for a fixed value of $d$. This  determines $\pi \ell $ for each $d$. We finally obtain the 
 plot in figure \nref{PlotLvsd}. For $d/d_0 \ll  1$ we have that $y$ is very large and we recover the approximation used in \nref{Vall} where $\ell \gg d$ and 
 $\pi \ell = d_0$. 
 In the other extreme, for $d/d_0 \gg 1$, we find that $\pi \ell $ grows linearly with $d$, with a coefficient which is greater than one, and numerically of order 2.35. 
 This ensures that in all cases $\pi \ell > d $, which is saying that it takes longer to go through the wormhole than through the ambient space. 
   In this regime the energy  gap is of order $E_{gap} \sim 1/d$, which 
   also the energy of a massless particle whose wavelength is the separation between the two black holes.   
 In the large distance regime the energy at the minimum depends on $d$ as  
    \be
  E_{\rm min} \propto  - { q \over  d }  ~,~~~~{\rm for}~~~~  q^2 l_ p \ll d 
 \ee
 This is larger than the gap by a factor of order  $q$.  
    
   Since the Landau levels are localized on the sphere, and the various levels have different lengths,  we have a Casimir energy that depends on the angle on the sphere. 
   So we  get a stress tensor which  depends on
   the sphere coordinates. This will excite the Kaluza Klein modes, which were found in 
   \cite{Michelson:1999kn}\footnote{In \cite{Michelson:1999kn} an extremal  black hole in supergravity was considered. However, the bosonic spectrum is the same as one in the Einstein Maxwell theory.}. Since Kaluza Klein modes are massive, this non-uniform energy distribution is not expected to be a problem, we will 
   simply slightly excite some of these Kaluza Klein modes. But we expect that the basic shape of the wormhole is still the same. The extra energy due to this 
   excitation of the KK modes is of higher order in the $1/q$ expansion compared to \nref{EnSu}. 
     
  It is interesting to ask whether there is an upper limit on $d$ for the validity of the solution. 
  We already mentioned that we expected that when $\ell \sim q^3$ quantum corrections would 
  be important \nref{QuanSure}. That estimate was based on an analogy with the thermal case. 
  Here we note the following, the circle on which the fermions live has a portion inside the wormhole throat and a portion outside, both proportional to $d$, up to a numerical factor. Though the total energy on the circle is fixed to be the Casimir energy, the energy within each portion could have quantum fluctuations. 
  We can worry that the total energy inside the throat might have fluctuations that are larger than
  the left hand side in \nref{Econ}, which would invalidate our classical derivation in \nref{MidEq} \nref{DetL}.  Estimating these fluctuations, after averaging them over a time also of order $d$, we find that they become larger than the left hand side of \nref{Econ} unless 
  \be
  \la{ConValQ} 
  d   \ll q^{5/2}
  \ee  which is smaller than our previous estimate $q^3$ in \nref{QuanSure}.
  The solution might exist beyond $d> q^{5/2}$, but it would require a more quantum mechanical treatment that we will not attempt in this paper.  
  The energy gap for throat excitations continues to be given by \nref{ThGap}. This is also true for energy of fluctuations of the $\ell$ mode around the minimum\footnote{In the $d\ll \ell $ regime the kinetic term for this mode comes from the Schwarzian action \cite{Maldacena:2018lmt}. 
  In the regime $d\gg \ell$, the dominant contribution to the kinetic term arises from the time dependent Casimir energy. Namely, if $L$ varies slowly, in addition to the Casimir energy, we get
  a term in the effective action of the form $S \propto  c \int du {{ L'}^2 \over L }$, for $L' = { d L \over du } \ll 1 $. }.

    
   \subsection{Some possible throat instabilities} 
   
   In a theory of quantum gravity we also expect magnetic  monopoles. These  monopoles could be pair  created in the throat resulting in a net decrease
   of the magnetic field in the throat\footnote{ The weak gravity conjecture implies that pair creation is always a possibility \cite{ArkaniHamed:2006dz}.}. 
   However, we expect that for large $q$ this process should be suppressed exponentially, as $e^{ - q \, C }$ where $C$  depends on the mass of the
   monopole, the Newton constant, the gauge coupling, etc, but not on the charge $q$. 
   
    Another possible issue is the following. If we have a single extremal black holes, we can split it into two different black holes with charges $q_{1,2}$, $q_1 + q_2 = q$ and separate them with no cost in energy. If we have a black hole/anti-black hole pair and we separate both the black hole and the anti-black hole in this way, then 
    we could imagine that we will form smaller throats. Since the binding energy of the throats goes like $-1/q$, we see that it is energetically favorable to split the
    black holes in this way. We expect that this is also exponentially suppressed for large charge. In fact, for $q_1=1$, this is a special case of the monopole 
    creation process discussed above.

 \section{Stabilizing the two wormhole mouths} 
 
 So far, we have only found a solution in the throat region.  We now need to pay attention to the exterior region, at distances $r\gg r_e$. 
 In this region,  we will treat the two mouths  as point masses.
  This is an approximation that has been extensively discussed, see eg. \cite{Goldberger:2004jt}. 
  We will approximate the two objects as extremal black holes, with opposite magnetic charges. 
These would attract and  would fall into each other if released at  rest.  So we need some mechanism to keep them apart. 
  
 \subsection{Rotation in flat space}

 The first mechanism involves putting the two  extremal black holes in orbit around each other at 
 a distance $d$ much larger than the size of each of the two black holes, $d \gg r_e$. 
 Then the orbital frequency for a circular orbit can be calculated by  the usual Kepler formula
 (the total force includes gravity and the magnetic attraction)
 \be
 \Omega = 2 \sqrt{ r_e \over d^3 }    \la{AngFre}
 \ee
 We would like $\Omega$ to be small compared to the energy gap of the throat region. 
 That energy gap is of order $1/\ell$. So we want 
 \be
 \Omega \ell \ll 1  \la{LowOm}
 \ee 
 This implies that $d$ should be sufficiently large.  In the regime that $ d\ll q^2$,  this implies
 \be
  l_p q^{5/3}  \propto   \left( { r_e^7 \over l_p^4 q^2 } \right)^{1/3}  \ll d   \la{FivTh}
 \ee
 where the first   term indicates the parametric dependence on $q$.
We see that, for large $q$ there is a whole
range of distances, $q^{5/3}  l_p \ll d \ll q^2 l_p  $, which are possible. 
In fact, the situation is even better, because for larger distances, $ q^2  l_p \lesssim d \ll q^{5/2} l_p $, then $ \ell \propto d $ and \nref{LowOm} is automatically obeyed. 
Therefore \nref{FivTh} is the lower bound on $d$. The upper bound is $d \ll q^{5/2} l_p$ from \nref{ConValQ}. 
 
 We could wonder whether we can tolerate a small excentricity in the orbit. This would result in perturbations of the throat with frequency $\Omega$. Since this is smaller than the energy gap in the throat, \nref{ThGap},  we expect that small excentricity should not destroy  the throat.

 \subsubsection{Effect of rotation on the fermions} 
 
 When the black holes are not rotating,  the lowest Landau levels are characterized by their azimuthal angular momentum, $m$. 
 This is a good quantum number both in the throat region and  the outside. 
 When the black holes are rotating this is no longer the case, so we could worry that the various modes are mixed with each other as they go through the exterior from 
 one mouth to the other. The fermions  will feel extra forces, coming both from the electric field generated by the rotation as well as the 
 Coriolis force.  Forces acting on  the lowest Landau level states lead to a transverse displacement:  the quantum Hall effect. This can be characterized as a geometric  angular  displacement of the orbitals as we go from one mouth to the 
 other. This is potentially dangerous because the electron starting in one field line can end on a different field line after going through one cycle, via  the wormhole and the exterior. This implies that it would have to go through many cycles before coming to the initial line. So it would be moving on an effectively larger circle, thus weakening
 the strength of the Casimir energy. 
Naively, the displacement of field lines is  proportional to the ratio of electric to magnetic fields, which is of order 
$\Omega d$. If we require that this is smaller than the effective 
  spacing between the 
 lines, $\Omega d \ll 1/\sqrt{q}$, we get a stringent condition $d \gg q^2 l_p$ which is only obeyed in the regime of section \nref{LargeDis}.  
 
 Fortunately, the massless charged fermions themselves largely screen the electric field in the rotating frame. 
 The residual electric fields, as well as the Coriolis terms are of order one in the $1/q$ expansion, see appendix \ref{ElectricFi}.
  This means that any possible drift of the fermions orbitals   is down by another power of $q$, so that now we need $\Omega d/q \ll 1/\sqrt{q}$ which is obeyed automatically for  $d> q l_p$.

 \subsubsection{Emission of radiation} 
 \la{Unruh} 
 
 The rotation also implies that we are emitting gravitational radiation as well as radiation of the gauge field (we can call it ``electromagnetic'' radiation). 
 The emission rates for each of these goes as 
 \be
 { d E_{\rm grav } \over d t } = { 2 \over 15 }{ r_e^2 d^4 \Omega^6 \over G_N } \propto { l_p^3  q^5 \over d^5} 
 ~,~~~~~~~~~~~ {d E_{\rm e.m. } \over dt } = \frac{4 \pi}{g^2} { 2 \over 3 } \lll \frac{qd}{2} \rrr^2 \Omega^4 \propto { l_p^2 q^4 \over d^4} 
 \ee
where we have given the $q$ dependence in the last terms. 
We can also wonder whether we can radiate fermions. We expect that this should be down by additional powers of $g^2$ since they need
 to be pair produced by the other fields.
 
  We see that the electromagnetic emission dominates. 
Due to this emission,  the total lifetime is finite 
 \be
 T_{\rm lifetime} \propto { d^3 \over q^2 l_p^2 } 
 \ee
which is small compared to the  time to go through the wormhole, when \nref{FivTh} is obeyed. 

The amount of energy 
 emitted per orbit is small compared to the gravitational binding energy of the system.  
     However, we might worry that some of this energy might make it into the throat. 
 We can think of the radiation as produced far away, in the region that is 
 at some distance of order $1/\Omega \gg \ell \gg d $ from the two mouths. 
 
   As a very strong condition, we could  demand  that the total energy emitted during a time of order $\ell$ is small compared to 
 $1/\ell$ we get the condition 
 \be   \la{EnTH}
 \ell  { d E_{e.m.} \over dt } \ll { 1 \over \ell }   \longrightarrow d \gg q^2 l_p 
 \ee
 This is obeyed in the regime describe in subsection \ref{LargeDis}. 
 However, this seems to be too stringent. We do not think we need it, since most of the radiated energy comes from the region far away 
 from the two black holes. If we demand that the left hand side of \nref{EnTH} 
  is small compared to the energy necessary to destroy the throat, which is $q/\ell$, then we see that
 it is automatically obeyed.

 \subsubsection{Unruh-like temperature } 
 
  One irreducible contribution to the energy that can go into the throat comes as follows.  
 
The fact that the two objects are accelerated implies that there is a an Unruh-like temperature.  This temperature is 
 of order the angular frequency $\Omega$ \nref{AngFre}, see e.g. \cite{Letaw:1979wy}. 
 One way to qualitatively understand this is to notice that the rotating configuration is invariant under the Killing vector 
 $\partial_t + \Omega \partial_\varphi $. This is timelike for small $r$ but becomes spacelike 
 for large enough $r$. The distance where it becomes spacelike, $\Omega \sim 1/r$, sets the 
 frequency of the radiation as well as the scale of the Unruh-like  temperature \cite{Letaw:1979wy}
 \be
 T \propto  \Omega 
 \ee
 (this formula  has multiplicative  logarithmic corrections in $\Omega d $ and the spectrum is not quite thermal). 
  Note that the temperature is {\it not} equal to the acceleration, $a$, which is much smaller  $a \sim \Omega^2 d \ll \Omega$ in our regime. 
 Fortunately,  \nref{LowOm} already ensures that this temperature is much smaller than the energy gap of the throat, which is of order $1/\ell$. If the temperature had been larger than the 
 this energy gap, we would have started filling the throat with these thermal excitations which would make the energy less negative, producing a longer wormhole, which in turn would be easier to fill with excitations, a process which will likely end up with the formation of a horizon. 
 Fortunately, this does not happend and  the Unruh-like temperature does not destabilize the throat.

 \subsubsection{Thermodynamic analysis for the rotating case} 
  \la{ThPot} 
  
 Given that the throat is effectively at finite temperature we can 
  ask whether it is the most stable phase in the canonical ensemble. 
 Another alternative configuration consists of two separate finite temperature black holes 
 at temperature $T \propto  \Omega$. In this case we lose the wormhole energy \nref{Vall} but we gain the entropy 
 of the extremal black holes, which is very large, $S\propto q^2$. 
Then the free energy of the two configurations is 
\be
F_{\rm wormhole} \propto  - E_{\rm min} ~,~~~~~~~~~F_{\rm 2BHs} \propto  - T  S \propto -\Omega q^2 
\ee
for $T \propto  \Omega$.  We   used that the extremal entropy is $S \propto q^2$, \nref{NeEn}.
The value of the wormhole energy is $E_{\rm min} \propto  -1/( q l_p) $ for $d< q^2 l_p$ and $E_{\rm min } \propto  - q/d $ for 
$d> q^2 l_p $.  
It turns out that for $d< q^3 l_p$ the wormhole is not the thermodynamically preffered state. It is 
metastable, and we expect it to be long lived. Analyzing it beyond $d> q^3 l_p $ requires  incorporating 
 the quantum gravity 
effects in the throat, which we are not doing in this paper.

 \subsection{Black holes in  $AdS_4$ } 
 
 If we are willing to put the system in $AdS_4$ then several options open up. 
 We just mention a few possibilities, leaving 
 further study for the future. 
 
 \subsubsection{Rotation in $AdS_4$} 
 
 In order to produce a more  stable wormhole we can put the whole rotating configuration in 
 $AdS_4$. Now we have a new parameter, the $AdS_4$ radius, $R_{AdS_4}$. If $d< R_{AdS_4}$ and 
 $\Omega R_{AdS_4} < 1$, then we can ensure that the symmetry of the rotating configuration 
 is described by a killing vector which is timelike everywhere. In this case, by weakly coupling 
 the $AdS_4$ theory to an external system we can let the black hole cool down and so that the thermodynamically 
 stable phase becomes the wormhole one.
 
 \subsubsection{Hovering black holes}

 When we are in $AdS_4$ we can fix the magnetic field at the boundary in such a way that 
 we produce a pair of black holes which sit at specific locations in the bulk. A configuration of this type consisting 
 of a single black hole is the magnetic dual of the solution  discussed in  
 \cite{Horowitz:2014gva}. 
  It seems possible to arrange the configuration so that a portion  of the magnetic field lines go between the two black holes. 
  
  \subsection{Two coupled $AdS_4$ spaces}
  
 Of course, we can also have two $AdS_4$ spaces, with one black hole at the center of each of the $AdS_4$ spaces. 
 We then couple the two $AdS_4$ spaces by putting boundary conditions for the fermions that allow them to 
  go from one $AdS_4$ to the other. This is simply a small variation of
 the kind of setup studied in \cite{Maldacena:2018lmt}.

 \section{Large number of flavors } 
   
 The parametric scalings we worked out in the main part of the paper apply when
we have a single bulk charged fermion. Instead we could have $N_f$ charged
fermions. 
Now the number of massless fields in the throat region 
 is $q N_f$ and the length and energy  of the wormhole scales as 
 \be \la{Flavors}
 \ell = 16 {r_e^3 \over G_N q N_{f} } \propto { q^2 \over N_f } l_p
~,~~~~~~~E_{\rm min} = - { G_N q^2 N_f^2 \over 256 r_e^3  } \propto- { N_f^2
\over q }{ 1 \over l_p}  ~,~~~~~~~~ q N_f \gg 1 \ee
where the last inequality ensures   the validity of the approximations, where we assume that the quantum correction is relatively small compared
to the classical action. 
In this case we can obtain wormholes which allow passage of objects smaller than $r_e$. 
This also allows for a range of distances, $d$, where the wormhole is the thermodynamically favored state, which was not the case for $N_f =1$, discussed
 in section \nref{ThPot}.

\section{Traversable wormholes in the Standard Model} 

Here we discuss how to embed the previous construction in the Standard Model with Einstein  gravity. 
We  (unconventionally) 
 normalize the weak  hypercharge so that the lowest charged fermion has charge one. 
We then consider a magnetic flux  
 $q$ which is an integer\footnote{ In principle, $q$ can be a multiple of $1/6$  at the cost of introducing fluxes of $SU(2)$ and/or $SU(3)$. We set $q $ 
 to an integer here.}. 
  We can imagine that $q$ is such that the radius of the extremal black hole, $r_e$, is much smaller than the electroweak scale (say 1/TeV). In this case, the Higgs field 
  in the throat is set to zero, since it is a charged boson in a magnetic field, whose lowest Landau level  has positive energy.

 We have a set of four dimensional positive chirality fermions of charges $q_i$.  The ones with positive $q_i$ lead to negative two dimensional chirality while the
  negative ones lead to positive two dimensional chirality. Note that the four dimensional antiparticle has opposite charge and opposite chirality, so that it gives rise
  to a mode with the same two dimensional chirality, but opposite charge, as expected for a two dimensional charged chiral fermion. 
   Each four dimensional fermion of charge $q_i$ leads to $q q_i$ massless fields in two dimensions, via
  the Landau level problem, since only this product appears in the equation for the fermions in the magnetic field\footnote{
  The fact that in the presence of fluxes we can get    massless 
  fermions   is a central element in Kaluza Klein compactifications. The idea that  weak hypercharge flux 
  can lead to light modes was introduced in Kaluza Klein theory in \cite{Donagi:2008kj}.  }. 
  Then the total central charge of the throat is 
  \be \la{CeCh}
  c = q   { 1 \over 2} \sum_i |q_i|
  \ee
  The mixed anomaly cancellation $\sum_i q_i =0$, ensures that there is the same number of positive and negative chirality modes, or $c_L = c_R$.

We have  the following charges for each generation, in units where the field with lowest hypercharge carries 
 hypercharge one. 
 \be \la{Chas}
 q_{e_R} = -6 
    ~,~~~~~~ q_{\, l_L^c} = 3   ~,~~~~~~~q_{u_R} =4  ~,~~~~~~q_{d_R}  = - 2 ~,~~~~~~q_{q_L^c} = - 1 
  \ee
  We view  all  them as  right-handed Weyl spinors, so we have conjugated the left-handed fermions. 
   Each of these comes with a degeneracy related to their $SU(2)\times SU(3)$ quantum numbers\footnote{In the order listed in \nref{Chas} the  degeneracies are 
   $(1,2,3,3,6)$. }. 
  Including the three generations, from \nref{Chas}
   we find that the net contribution to the central charge is the same as if we had an {\it effective}  number of
  flavors $N_f = 54$ of charge one\footnote{ 
   Alternatively, we keep the usual 
 hypercharge assignments, say that $q$ is a multiple of six and an $N_f = 9$. }, in formula \nref{Flavors}. 
 
 In order for the analysis in the previous sections to be valid, we need that the distance $d$ between the two black holes is smaller than the electroweak scale, so 
 that we can approximate all the fermions as being massless. Since we had a hierarchical relation between $d$ and $r_e$, we see that the size of the black holes 
 would be much smaller than the electroweak scale. The coupling $g$ that appeared in \nref{Metr} is the hypercharge coupling evaluated at the scale of the magnetic
 field in the throat region. 
 
 One could imagine also a configuration where the distance $d$ is smaller than the Compton wavelength of the electron:  $1/m_e$. In this situation we can imagine that the 
 two objects carry ordinary magnetic field. However, in this case we expect that $r_e$ should be much smaller than the electroweak scale in the throat region. 
 Then the $SU(2)$ component of the magnetic field would be screened in the throat\footnote{ Depending on whether the magnetic field flux is even or odd, we could 
 be left with a discrete flux of the $SU(2)$ gauge field, since this is small relative to $q\gg 1$, we will not worry about it.}. So we expect that near the mouth of 
 each black hole we will have a  condensate of $W$ and $Z$ bosons. The Higgs field also would go from the non-zero value in the vacuum to zero inside the throat. 
 This structure would be present for any magnetically charged black hole  with radius, $r_e$,  smaller than the electroweak scale.
  Here in addition we are connecting the two throats. It would be interesting
 to find this configuration in detail. Large magnetic fields were discussed in e.g. \cite{Ambjorn:1992ca}.  
 
 Of course, since we are talking about distances shorter than the ones experimentally measured, it could be that the Standard Model is not a good description. 
 For example, there could be other particles carrying weak hypercharge, etc. 
 
 New physics beyond the Standard Model could also give rise to new opportunities to construct this solution. There could be light dark photons, for example. 
 Also in a Randall-Sundrum model without an IR brane (RS2) \cite{Randall:1999vf}, 
 which is equivalent to coupling the SM to a CFT, we could use these new light degrees of freedom 
 as the ones propagating inside the throat.  
  This raises an interesting question. In a solution of this kind, where the matter has an AdS dual,   then quantum effects become classical. This means that such a solution 
  should not be classically forbidden by the topological censorship arguments \cite{Galloway:1999bp,Galloway:1999br}.  We think that this means that the throat region should be continuously connected with the exterior via the 5 dimensional bulk. 
  It would be nice to see whether  a solution of that form is possible. 
   
 \section{Conclusions and discussion }

 \subsection{Summary} 

 We have constructed a traversable wormhole solution in four dimensions. 
 It is a solution of a theory described by Einstein gravity plus a $U(1)$ gauge field plus massless charged fermions.
 Solutions with this topology are forbidden in classical physics because of the average null energy condition, which is positive classically. The idea is to engineer a configuration that
 generates enough negative null energy via quantum effects so as to make the solution possible. This is achieved by   close to  extremal black holes  with a large magnetic charge, 
  $q\gg 1$. 
 Charged fermions moving in the magnetic field have a zero energy Landau level on the sphere with  large degeneracy proportional to $q$. These give rise to $q$ two dimensional fields moving 
 along the magnetic field lines. In the wormhole configuration these magnetic field lines describe closed circles. A massless fermion on a circle develops a negative energy. This negative energy, when fed into Einstein's equations, makes the wormhole into a solution. 
 To stabilize the mouths,  we added rotation so that they can be kept apart, at least for a while until the electromagnetic radiation causes them to lose energy and eventually merge with each other. 

We should emphasize that this wormhole does {\it not} require any exotic matter. In fact, the ordinary matter of the Standard Model is enough!.

 The wormhole is metastable, it is a long lived solution, but it is not completely stable for 
 various reasons. 
 
 The resulting wormhole solution has no horizon and no entropy. It can be viewed as an 
 entangled state of two black holes. The state is close to the thermofield double. If there was 
 no interaction between the two black holes, the thermofield double state would be a time dependent state. The interactions change the Hamiltonian and now the thermofield double is 
 close to the true ground state of the combined system. This is explained in more detail in 
 \cite{Maldacena:2018lmt}, and results from a mechanism related to the one proposed in \cite{Gao:2016bin}.
  
  It might seem surprising that we get a controllable solution by balancing classical and quantum effects. There are a couple of features that make this possible. 
  First, in the presence of a large magnetic flux, $q\gg 1$, we end up with a large number of massless two dimensional fermions, of order $q$. It is important that
  we get this large number of light fields both in the throat as well as in the flat space region between the two mouths. 
  The second  feature is the appearance of a scaling regime as we approach extremality, where the black hole develops a long throat. 
  This long throat is ``almost'' a wormhole, in the following  sense.  The global $AdS_2 \times S^2 $ geometry looks like a wormhole.
   However, when we deform one of its   ends 
  so that it joints on to flat space the classical equations produce a singularity in the other. More precisely, when the sphere expands at one end to lead to flat space, it
  shrinks at the other producing a singularity. The  negative null-null components of the stress tensor make it possible to 
  make the two spheres grow as we go to both boundaries. 
  The classical action of the solution is of order $q^2$, and the effect of these fermions is of order $q$. The length, or depth, of the throat is very large, 
  larger than $q^2 l_p$.
  The classical action is of order $q^2$ and the Casimir energy is of order $q/l_p$.  The large length of the wormhole reduces the classical action of one particular mode, 
  the boundary graviton of $AdS_2$, so that we can balance this term against the Casimir energy  \cite{Maldacena:2018lmt}. 
   The presence of this long throat also removes the quantum effects of all other fields that  are massive in   the $AdS_2$ sense. More precisely, 
  the quantum effects are there, but they do not change the fact that we get an $AdS_2$ solution. 
  
  Notice that the proper time that it takes for an observer to go through the wormhole is of order the light crossing time of the black hole, or $r_E \sim q $. This is much smaller than the time it takes to go through the wormhole as seen from the outside, which is $\pi \ell \propto q^2$. 
  
  \subsection{Van der Waals analogy} 
  
  The fact
  that close proximity leads to entanglement is a very common
   occurrence for systems that have several low energy levels. A simple example is given by the Van der Waals force between
  two neutral atoms. This is produced by the exchange of photons, when the distance between the atoms is larger than their size but  smaller than the inverse energy spacing of the atomic levels. In this case, after integrating out the photon, we induce a dipole-dipole interaction of
  the schematic form $H_{\rm int} \propto  { \vec p_L . \vec p_R / d^3} $ where $d$ is the distance, and 
  $\vec p_{L,R}$ are the  dipole operators of two atoms, see e.g. \cite{Brambilla:2017ffe}. This interaction produces entanglement among the low energy levels of the atom and a potential which goes as the square of this 
  interaction Hamiltonian,    $V \propto - 1/d^6$. 

In our system, instead of atoms we have two black holes, with an exponentially large number 
of energy levels. When they are in close proximity they exchange the charged fermions (and other fields). This generates an effective Hamiltonian of the form $H_{\rm int} \propto  \psi_L^i \psi_R^i $ acting  in the near horizon regions of the two black holes. The ground state of this Hamiltonian is then given by the wormhole configuration. This is precisely the kind of
Hamiltonian studied in \cite{Maldacena:2018lmt}. 
In  \cite{Maldacena:2018lmt}, this interaction was introduced by hand. Here we see how it arises naturally when two black holes are close to each other. 
 In principle, we also generate terms in the interaction Hamiltonian 
involving other fields. But since the other fields are  massive in the throat region, they 
do not generate much negative energy.

\subsection{Open questions} 

In this paper we showed that the solution exists, but we did not describe a simple procedure for
taking the system to that solution.  Ideally, one would like to be able to give a procedure 
to construct the solution that is not exponentially difficult as the charge becomes large. 
For example, one would like to say that one could produce many monopoles, then get them 
to collapse into two separate black holes, then let them evaporate close to extremality, then bring them close enough so that the wormhole can form. It is not clear how long this last step 
would take, or how to steer the solution into the wormhole rather than the thermodynamically preferred solution consisting of two separate black holes. 

The mechanism by which the wormhole forms is bound to be very interesting, since it involves topology chance and the disappearance of horizons. Of course, we already know that such processes can occur by tunneling, with exponentially small probabilities. There are some reasons to believe that in coupled SYK models, such as the one in \cite{Maldacena:2018lmt}, this process occurs 
in a time which is a power law in $N$, the size of the system. But we do not know 
whether this is also the case for these Einstein gravity wormholes. 
    
 Understanding these questions would  give us more insight into  the generation of spacetime connections from entanglement. 
 More optimistically, it might lead to some futuristic experiment for checking that this connection indeed appears.

 It would also be interesting to see whether a similar construction can be done in a supersymmetric or BPS context.

{\bf Acknowledgements } 

We thank A.~Almheiri, K.~Bulycheva, D.~Harlow,   S.~Knapen, D.~Marolf, X.~Qi,  N.~Seiberg,  D.~Stanford, C.~Vafa,  A.~Wall, E.~Witten 
and  Z.~Yang for discussions. 
J.M. is supported in part by U.S. Department of Energy grant
de-sc0009988 and the It from Qubit grant from the Simons foundation. F.P. is supported in part by the U.S. NSF under Grant No. PHY-1620059.

\appendix 
\setcounter{equation}{0}
  
\section{Mass generation}

  \la{MassGen}
    
   We have shown that a fermion  field  on  $AdS_2\times S^2$  leads to $q$ two dimensional massless fermions. 
   Now we will take into account the dynamics of the gauge field. Of particular interest is the dynamics of the two dimensional gauge field 
   along the $AdS_2$ directions. 
   
   First let us consider a two dimensional theory in flat space consisting of $n$ charged fermion fields interacting with a two dimensional gauge field 
   \be
   S = \int d^2x  \left[ - { 1 \over 4 g_2^2 } F^2 +  i \sum_{l=1}^n  \bar \chi_l \gamma^\alpha ( \partial_\alpha - i A_{\alpha } ) \chi_l \right]
   \ee
   We can bosonize each of the fermion fields into a boson $\varphi_l$. The gauge field couples to the ``center of mass'' $\varphi={1 \over n} \sum_{l=1}^n \varphi_l$ only.
   The Lagrangian for this mode then becomes 
   \be
   S = \int d^2 x \left[ - { 1 \over 4 g_2^2 } F^2 +  {n \over 2 \pi } (d \varphi - A)^2   \right]
   \ee
  Fixing the gauge $\varphi=0$,  we see that the gauge field describes a massive mode of mass $m^2 =  g_2^2 n/\pi $. 
   The rest of the modes remain massless. This can be argued as follows. We can bosonize the fermions in terms of the overall $U(1)$ degree of freedom $\varphi$ and
   an $SU(n)_1$ WZW model. The fact that this $SU(n) $ chiral symmetry (left and right) 
    is preserved by the quantization implies that we should continue to have massless degrees of freedom. In this case they are described by the $SU(n)_1$ WZW model. 
    
    If we now go back to the $AdS_2 \times S^2$ case, then we find a two dimensional gauge field, with an action of the form 
    \be
    S = { 4 \pi \over 2 g^2 } \int dt d\sigma \sin^2 \sigma F_{t\sigma}^2  + \sum_m \int dt d\sigma i \bar \chi_m  (\slashed{\partial} - i \slashed{A})  \chi_m 
    \ee
    where the $AdS_2$ metric is $ds^2 = ( -dt^2 + d\sigma^2)/\sin^2 \sigma$. We see that we get very similar physics so again we can bosonize and obtain 
    a massive mode, with mass $m^2R_{AdS_2}^2 \propto g^2 q $, where $g$ is the four dimensional coupling of the gauge field.  
    This is only one mode, and the rest of the modes remain massless. More precisely, the other Kaluza Klein modes of the gauge field on the sphere
    also get an additional contribution to their (mass)$^2$, of the form $g^2 q$, 
       due to the interaction with the massless fermions. However, since the initial vector field was already massive, 
    we are still left with   massless modes. 
    
    \subsection{Non abelian gauge fields } 
    
    We can also consider  a non-abelian $SU(r)$ gauge field interacting with $n$ fermions in the fundamental representation. 
    In this case the non-abelian bosonization will give us an overall $U(1)$ scalar and an $SU(r)_n$ and an $SU(n)_r$ WZW model. 
    The coupling to the gauge fields eats the $SU(r)_n$ degrees of freedom. But we still remain with the $SU(n)_r$ WZW model with a number of 
    degrees of freedom which is of order $n r$ for large $n$ and fixed $r$. This is simply because, when we gauged the $SU(r)$, we only removed of order $r^2$ 
    degrees of freedom. 
    
    In the case that we consider the Standard Model, we are in this situation. We are gauging the $U(1) \times SU(2) \times SU(3)$ symmetries and this removes
    a number of massless degrees of freedom. However, the number we remove is fixed when $q$ becomes large. 
       
       We can also consider the $SU(2)$ Kaluza Klein
        symmetry that acts on the sphere. At leading order,  these give rise to massless two dimensional gauge fields. The fermions are charged under these gauge fields since they transform in the spin $j =(q-1)/2$ representation of $SU(2)$. This gives the gauge fields  a mass of  order $m^2 R_{AdS_2}^2 \propto
         { l_p^2 q^3 /r_e^2 }\propto  g^2 q$. This is curiously of the same order as that of the $U(1)$ boson.

  \section{The next order correction in the gravity solution} 
  \la{gamma} 
   
   In this appendix we discuss the form of $\gamma $ in \nref{FdAn}. We look at the sphere components of the Einstein equations and get the following
   equation for $\gamma$ 
   \be
    \gamma''   + ( 1 + \rho^2 ) \phi'' + 2 \rho \phi'  + 4 \phi =0 
   \ee
   Integrating this equation we get
   \be
   \gamma = \tilde c  - \alpha \left[  \rho^2 + \rho ( 3 +  \rho^2) \arctan \rho - \log(1 + \rho^2 ) \right] 
   \ee
   where we have imposed the $\rho \to - \rho$ symmetry to fix one of the integration constants. $\alpha $ is given in \nref{MidEq}. 
   The constant $\tilde c$ can be set to zero by 
   a rescaling of the $\rho$ and $\tau$ coordinates. So we set it to zero. 
   It is interesting to expand $\gamma$ for large $\rho$. 
   We obtain 
   \be \la{gamE}
   \gamma \sim \alpha \left[ - { \pi   \over 2 } \rho^3 - { 3  \pi \over 2 } \rho + 2   \log \rho + \cdots   \right] 
\ee
We can now do the same comparison we did in \nref{gttCom} but including $\gamma$ and expanding to the next order in $r-r_e$ in \nref{ABva}. 
 We then find that the leading order term in $\gamma$ \nref{gamE} matches the third  term in the $r-r_e$ expansion of \nref{ABva} 
 \be \la{Aexpt}
 A = { (r -r_e)^2 \over r_e^2 }  - { 2 \varepsilon \over r_e}  - 2  { (r-r_e)^3 \over r_e^3}     + { 2 \varepsilon \over r_e} { (r -r_e) \over r_e }  -  { 2\varepsilon \over r_e }    { (r-r_e)^2 \over r_e^2} + o( (r-r_e)^4)
\ee
This is to be compared to the expansion of 
\be  \la{CubE}
 r_e^2 ( 1 + \rho^2 + \gamma) { d\tau^2 \over dt^2 } = { r_e^2 \over \ell^2 }  ( 1 + \rho^2   - { \pi \alpha \over 2 } \rho^3 - { 3 \alpha \pi \over 2 } \rho + \cdots ) 
\ee
 Note that $\epsilon \propto \alpha^2$ and $r-r_e$ should be viewed as being of order $\alpha$, in light of \nref{rvsrho}.
Using \nref{rvsrho} and \nref{DetL} we see have already checked that the first two terms match in \nref{gttCom}. 
Unfortunately, in order to match the next term we also need the next order solution in $\phi$. 

However, we can go to a regime where $\rho$ is large, but $\alpha \rho$ is small and  fixed, so that $\phi$ is still given by the first
order expression \nref{rvsrho}. This amounts to expanding around small $(r-r_e)/r_e$ but very small $\alpha$. Since $\epsilon \propto \alpha^2$, in this 
regime we can expand  \nref{Aexpt} as 
\be
A \sim  { (r -r_e)^2 \over r_e^2 }   - 2  { (r-r_e)^3 \over r_e^3}  + \cdots 
\ee
The cubic term then matches the cubic term in $\rho$ in \nref{CubE} coming from the cubic term in \nref{gamE}, using  \nref{rvsrho}.
 
\section{The solution of the Dirac equation }
We consider a Dirac equation on the following background gravitational and electromagnetic fields
\be
ds^2 = e^{2\sigma(t,x)}\left(-dt^2 + dx^2\right) + R^2(t,x) \left(d\theta^2 + \sin^2 \theta d\phi^2\right), \quad A = \frac{q}{2} \cos\theta d\phi
\ee
In this metric we can choose a simple vierbein
\be
e^1 = e^\sigma dt,\quad e^2 = e^\sigma dx,\quad e^3 = R d\theta,\quad e^4= R \sin \theta d\phi.
\ee
The only non-zero spin-connections are
\be
\omega^{12} = \sigma' dt + \dot \sigma dx, \quad \omega^{32} = R' e^{-\sigma} d\theta, \quad \omega^{42} = R' \sin \theta e^{-\sigma} d\phi, \quad \omega^{43} = \cos\theta d\phi,
\ee
where a dot corresponds to the differentiation with respect to the time and a prime with respect  to the spatial coordinate. 
We write  the four dimensional spinors as tensor product of two dimensional spinors $\chi_{\alpha\beta} =\psi_\alpha \otimes \eta_\beta$. The gamma matrices in this representations are
\be
\gamma^1 = i\sigma_x \otimes I, \quad \gamma^2 = \sigma_y \otimes I, \quad \gamma^3=\sigma_z \otimes \sigma_x,\quad \gamma^4 = \sigma_z \otimes \sigma_y.
\ee  
The Dirac operator in this representation is
\begin{gather}
\slashed D = e^{-\sigma} \left[i\sigma_x \left(\partial_t + \frac12 \dot \sigma\right) + \sigma_y \left(\partial_x + \frac12 \rho' + \frac{R'}{R}\right)\right] \otimes I +\notag\\+ \frac{\sigma_z}{R} \otimes \left[\sigma_y \frac{\partial_\phi - i A_\phi}{\sin \theta} + \sigma_x\left(\partial_\theta + \frac12 \cot\theta\right) \right]
\end{gather}
We substitute the following ansatz in the Dirac equation
\be
\chi_{\alpha\beta} = \frac{e^{-\frac12 \sigma}}{R} \psi_\alpha(x,t) \eta_\beta(\theta,\phi),
\ee
We now look for a solution where the $\eta$ spinor obeys 
\begin{gather}
\left[\sigma_y \frac{\partial_\phi - i A_\phi}{\sin \theta} + \sigma_x\left(\partial_\theta + \frac12 \cot\theta\right) \right] \eta=0, \\
  \left(i\sigma_x \partial_t+ \sigma_y \partial_x\right)\psi = 0
\end{gather}
The second equation indicates that we get an effectively massless two-dimensional spinor $\chi$.   
 The second equations can be exactly solved. For $q>0$ we get 
\begin{gather}
\eta_- = \left(\sin \frac{\theta}{2}\right)^{j-m} \left(\cos \frac{\theta}{2}\right)^{j+m} e^{i m \phi}, ~~~~\eta_+=0, ~~ \quad j=\frac{|q|-1}{2},\quad -j\le m \le j
\end{gather}
where $\sigma_z \eta_\pm = \pm \eta_\pm$. 
If $q<0$ we should switch $\eta_+ \leftrightarrow \eta_-$ and if $q=0$ there is no zero modes on the sphere.

Here we assumed that the four dimensional field has charge one. If instead it has charge $q_i$, then we should replace $q \to q q_i$ in the above formulas. 

If we started with a four dimensional Weyl spinor of a definite chirality, then, since $\eta$ has definite chirality, given by the sign of $q_i q$, then the spinor
$\psi$ also has a definite chirality. 

\section{Properties of magnetic field lines}
\label{flines}

\begin{figure}[h]
	\begin{center}
		\includegraphics[scale=0.7]{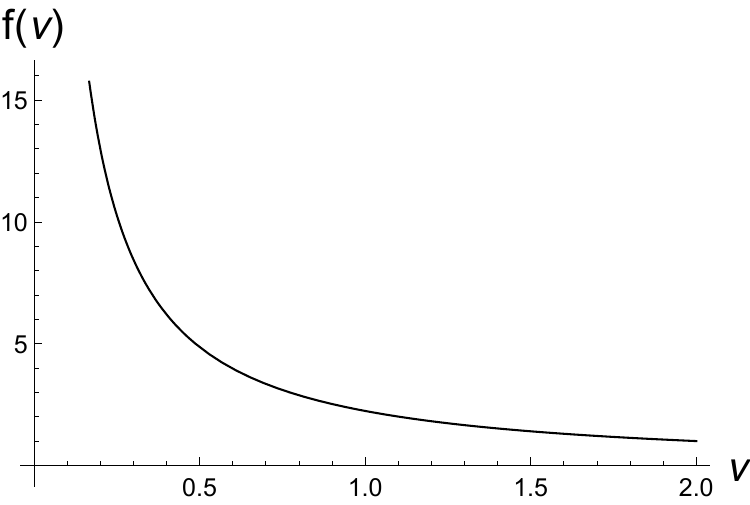}
		\caption{The   length of field lines as a function of $\nu$, in units of $d$. }
		\label{Plotlv}
	\end{center}
\end{figure}

 We now compute the length of the field lines in the flat space region in terms of $\nu$ in \nref{MFline}. 
We make a change of the coordinates from the coordinates $z,\rho$ to $\theta_1,\theta_2$ (see figure \ref{FieldLines})
\be
{ z \over d } = \hat z(\theta_1,\theta_2) =   \frac{\sin(\theta_1 + \theta_2)}{2\sin(\theta_2-\theta_1)}, \quad ~~~~~~~ { \rho \over d } =  \hat \rho(\theta_1,\theta_2) = d \frac{\sin(\theta_1) \sin(\theta_2)}{\sin(\theta_2-\theta_1)}
\ee
With the use of the equations for the field line \eqref{MFline} we can express one of the angles $\theta_2$ via the other
\be
\theta_2(\theta_1) = \arccos(\cos\theta_1-\nu),\quad ~~~~~~\theta_1 \in \left[0,\arccos(\nu-1)\right].
\ee 
After that the length can be calculated as
\be \la{Deffnu}
f(\nu) = { L  (\nu) \over d} = \int\limits^{\arccos(\nu-1)}_0 d\theta_1 \sqrt{(\hat z'(\theta_1,\theta_2(\theta_1)))^2+( \hat \rho'(\theta_1,\theta_2(\theta_1)))^2 }
\ee
The plot of this length is depicted on the picture \ref{Plotlv}

\section{Motion along the field lines}
\label{potential}
In this section it is  convenient to make the angular momentum $\nu q/2$ go between 0 and $q$, such that 
$\nu q=0$ corresponds to fermions going away from the black holes, and
$\nu q/2=q$ to fermions going straight between them.
The square of the Dirac operator is 
\beq
\slashed{D}^2 = -E^2 + (p - A)^2 + \frac{i}{2} F^{\mu \nu} \gamma_{\mu\nu}
\eeq
Lets ignore the spin term for a while. Then we get an effective Schr\"odinger equation:
\beq
E^2 = -\pr_z^2  -\pr_\rho^2 + \frac{1}{\rho^2}(i\pr_\phi + A_\phi)^2
\eeq
where we wrote it in cylindrical coordinates: $ds^2 = -dt^2 + dz^2 + d\rho^2 + \rho^2 d\phi^2$.
If the wave function $\psi$ has a definite momentum $\nu q/2$ in the $\phi$ direction then we get
\beq
\label{sch}
E^2 = -\pr_z^2 -\pr_\rho^2 + \frac{1}{\rho^2} \lll -{\nu q \over 2} + A_\phi \rrr^2
\eeq
For two monopoles $A_\phi = {q \over 2} (\cos \th_1 - \cos \th_2)$.
So we get an effective two-dimensional motion in the potential:
\beq
V = \frac{1}{4\rho^2} \lll \nu q - q \cos \th_1 + q \cos \th_2 \rrr^2
\eeq
This potential has a minimum along the field line and is separated by a barrier from infinity.
To estimate when a fermion can tunnel to infinity lets study the motion along the $\rho$ direction in this potential.
For simplicity consider the plane $z=d/2$(the middle plane between monopoles). Then:
\beq
V(\rho) = \frac{q^2}{4\rho^2} \lll \frac{d}{\sqrt{\rho^2+d^2/4}} - \nu \rrr^2
\eeq
Suppose that $\nu \ll 1$. Then the potential minimum occurs at $\rho\sim { d \over \nu }  $ and the maximum at $\rho \sim  2d/\nu$. 
Note that the barrier length is of order $d/\nu$, we will need it later.
To understand whether a fermion can tunnel or not we will compare the barrier height with the ground state energy near
minimum. We can estimate this energy by approximating the potential there by a quadratic oscillator. It is easy to obtain that:
\beq
\begin{split}
E_{bound} = \sqrt{\frac{V''}{2}} = \frac{(\nu q)^3}{2 q^2 d^2} \\
V_{max} = \frac{(\nu q)^4}{64 q^2 d^2}
\end{split}
\eeq
They are comparable only for $\nu q \sim 1$. It means that fermions with $\nu \gg 1/q$ will follow the field lines.

\section{Extra contributions to the two dimensional stress tensor due to the conformal anomaly } 

\la{ConfAn}

We consider a metric $ds^2 = e^{2 \omega } ( -dt^2 + dx^2 )$. 
 More generally, we can consider two metrics related by $g_{\mu \nu} = e^{ 2\omega } \hat g_{\mu \nu }$. 
 There are correspondingly two stress tensors $T_{\mu \nu} $ and $\hat T_{\mu \nu} $ obtained by taking a derivative with respect to each of the metrics. 
 The two are not the same because they are defined with different regularizations.
  The conformal anomaly allows us to relate the two partition functions. In Lorentzian signature we have
\be
 Z[ g = e^{ 2 \omega } \hat g ] =  \exp\left\{ i  { c \over 2 4 \pi  } \int  d^2 x  \sqrt{\hat g } [ \hat R  \omega +  ( \hat \nabla \omega )^2 ] \right\}  Z[ \hat g ] 
 \ee
 The stress tensors are related as 
 \bea \la{That}
 T_{\mu \nu}  & =&  \hat  T_{\mu \nu} -  { c \over 12 \pi  }   \left[ 
  \partial_\mu \omega  \partial_\nu \omega - \half \hat g_{\mu \nu } (  \hat \nabla \omega )^2 -  \hat  \nabla_\nu \hat \nabla_\mu \omega +
   \hat g_{\mu \nu}   \hat \nabla^2 \omega
\right] 
\cr 
 T_{\mu \nu} &=& \hat  T_{\mu \nu}  + { c \over 24 \pi } \left( \begin{array}{cc} 1 & 0 \\ 0 & 1 \end{array} \right)_{\mu \nu}  - { c   \over 24 \pi} 
 g_{\mu \nu} 
\eea
where we used $e^{ 2 \omega } \propto { 1 \over \sin^2 x }$.  The last term gives the contribution to the trace anomaly $T^{ \, \mu}_{\mu } = { c \over 24 \pi  } R $. 
But we also have the second term, which comes also from the conformal anomaly but is a traceless contribution. For standard $AdS_2$, with Dirichlet (or Neumann) boundary conditions for the fields, this term cancels precisely the vacuum energy in the flat space strip that we get in  $\hat T_{\mu \nu}$.  This is good because the 
stress tensor $T_{\mu \nu}$ should be $SL(2)$ invariant in that case, which implies that only a term proportional to the metric is allowed. 

In our problem, we have other boundary conditions and $\hat T_{\mu \nu}$ is the naive Casimir energy in the flat cylinder. Then  the second term in \nref{That} 
gives an extra contribution, which we were calling the ``conformal anomaly contribution'' in the main text.

\section{Electric fields for the rotating configuration } 

\la{ElectricFi}

In this appendix we discuss in some detail the electric fields that arise when we consider the two rotating magnetic charges. 

Let us assume that we  rotate  along the $\hat y$ axis. Then in the rotating frame we have the metric 
 \be \la{RotMe}
 ds^2 = - dt^2 + dx^2 + dy^2 + dz^2 + 2 \Omega dt ( z dx - x dz ) + o(\Omega^2 ) 
 \ee
 where we neglect terms of higher order in $\Omega$. 
 Let us also assume that the two magnetic charges are separated along the $z$ axis, at $z   = \pm d/2$. 
 To leading order we have the same magnetic field we had in the static frame, given by \nref{VecPo}. 
 However, due to the new term in the metric \nref{RotMe} this field no longer obeys Maxwell's equations to leading order in $\Omega$. 
 This can be fixed by adding a time independent non-zero component $A_t$ for the potential. So that now we have the Maxwell equation
 \be \la{MaxHom}
 \nabla_\mu F^{\mu t } = \partial^2 A_t - 2 \vec \Omega . \vec B =0  ~~~\longrightarrow A_t ={ q  \Omega y \over 2 } \left[ { 1\over |\vec r - \vec d/2| } - { 1 \over |\vec r + \vec d/2| } \right] 
 \ee
 to first order in $\Omega$. 
 Here $\nabla$ is the covariant derivative with the metric \nref{RotMe} and $\partial^2$ is the flat space laplacian in the spatial directions. 
 This is the naive electric field that we obtain if we ignore the fermions. Of course,  this electric field can be obtained directly from the usual 
 Lienard Weichert potentials,  applying an electric-magnetic duality transformation, and  going to the rotating frame.
  
 Let us now include the fermions. Most of the fermion modes in this region have large energies 
 and we can ignore them. We are left with the massless modes that 
 propagate along the field lines. These modes interact with the electric field. It is convenient to bosonize them so that along each field line we have an action 
 of the form 
 \be \la{BosAc} 
 S = { 1 \over 2 \pi } \int dt ds  \left[ (\partial_t \varphi - A_t)^2 -  ( \partial_s \varphi - A_s)^2 \right] 
 \ee
  where $s$ is the length coordinate along the field line  and $A_s = A_ i { d x^i \over d s }$. 
 In our case $A_s =0$ and we postulate an $A_t$ which depends only on space. In this case, the minimum energy solution of 
 \nref{BosAc} is one where $\varphi=$constant. For this solution we have a contribution to the electric current that goes as 
 \be
 j^t =    { 1 \over \pi }  A_t \delta^2_\perp 
  \ee
   where the delta function  is in the two spatial directions  orthogonal to the field line. This comes from the derivative of  \nref{BosAc} with respect to $A_t$ for $\varphi =$constant. 
   We  have a large density of field lines  going as $|B|/(2 \pi)$. After  we take into account all field lines we get 
   \be
   j^t = { |B| \over 2  \pi^2 } A_t 
   \ee
   Therefore, after including the fermions,   the Maxwell equation reads (to first order in $\Omega$)
   \be
   \partial^2 A_t - 2 \vec \Omega . \vec B =  \nabla_\mu F^{\mu t } =g^2  j^t =  { g^2 |B| \over 2 \pi^2 } A_t \longrightarrow -\partial^2 A_t + { g^2 |B| \over 2 \pi^2} A_t = - 2 \vec \Omega . \vec B 
   \ee
   with the rest of the components automatically obeyed. 
   The terms with an explicitly $B$ field factor   are very large for large $q$. This means that we can ignore the laplacian term and 
   approximate the solution as 
  \be
  A_t \sim - { 4 \pi^2  \vec \Omega . \vec B \over g^2 |B| } + o (1/q) 
  \ee
  This means that the electric field is of order one, rather than of order $q$ as in \nref{MaxHom}. 
  Therefore we find that the leading order electric field is screened by the charged fermions. 
  
  Therefore, in the Dirac equation all extra terms are of order one: the electric field and the off diagonal metric components (or Coriolis force). 
 The drift they induce is  of order the ratio of these new forces and the magnetic field, or 
  \be
   \delta \theta \propto  { \Omega d \over q } 
  \ee
  This is the drift   experienced by a fermion  going along 
   one magnetic field line. It starts at  some angle $\theta$ and it ends at the angle $\theta + \delta \theta$.

\mciteSetMidEndSepPunct{}{\ifmciteBstWouldAddEndPunct.\else\fi}{\relax}

\bibliographystyle{utphys}
\bibliography{GeneralBiblio}{}

\end{document}